\documentclass[twocolumn]{aastex61}
\usepackage{graphicx}
\graphicspath{{./Figures/}} 
\usepackage{float}
\usepackage{longtable}
\pdfoutput=1

\definecolor{emerald}{rgb}{0.27, 0.62, 0.14}
\definecolor{twilightsky}{rgb}{0.42, 0.77, 0.22}

\shorttitle{Modelling of the hydrogen Lyman lines in solar flares}
\shortauthors{Brown et al.}
\begin{document}

\title{Modelling of the hydrogen Lyman lines in solar flares}
\author{Stephen A. Brown}
\affiliation{SUPA School of Physics \& Astronomy, University of Glasgow, Glasgow, Scotland, G12 8QQ}

\author{Lyndsay Fletcher}
\affiliation{SUPA School of Physics \& Astronomy, University of Glasgow, Glasgow, Scotland, G12 8QQ}

\author{Graham S. Kerr}
\affiliation{NASA Goddard Space Flight Center, Heliophysics Sciences Division, Code 671, 8800 Greenbelt Rd., Greenbelt, MD 20771, USA}

\author{Nicolas Labrosse}
\affiliation{SUPA School of Physics \& Astronomy, University of Glasgow, Glasgow, Scotland, G12 8QQ}

\author{Adam F. Kowalski}
\affiliation{Astrophysical \& Planetary Sciences, University of Colorado, Boulder, USA, CO 80309}

\author{Jaime De La Cruz Rodr\'{i}guez}
\affiliation{Department of Astronomy, Stockholm University, AlbaNova University center, Stockholm, Sweden, SE-106 91}

\correspondingauthor{Stephen A. Brown}
\email{s.brown.6@research.gla.ac.uk}
	
\begin{abstract}
	The hydrogen Lyman lines ($ 91.2$ nm $< \lambda < 121.6$ nm) are significant contributors to the radiative losses of the solar chromosphere, and are enhanced during flares. We have shown previously that the Lyman lines observed by the Extreme Ultraviolet Variability instrument onboard the Solar Dynamics Observatory exhibit Doppler motions equivalent to speeds on the order of 30 km s$^{-1}$. But contrary to expectation, no dominant flow direction was observed, with both redshifts and blueshifts present. To understand the formation of the Lyman lines, particularly their Doppler motions, we have used the radiative hydrodynamic code, RADYN, and the radiative transfer code, RH, to simulate the evolution of the flaring chromosphere and the response of the Lyman lines during solar flares. We find that upflows in the simulated atmospheres lead to blueshifts in the line cores, which exhibit central reversals. We then model the effects of the instrument on the profiles using the EVE instrument's properties. What may be interpreted as downflows (redshifted emission) in the lines after they have been convolved with the instrumental line profile may not necessarily correspond to actual downflows. Dynamic features in the atmosphere can introduce complex features in the line profiles which will not be detected by instruments with the spectral resolution of EVE, but which leave more of a signature at the resolution of the Spectral Investigation of the Coronal Environment (SPICE) instrument on Solar Orbiter.
\end{abstract}
	
\section{Introduction} \label{sec:intro}
The solar chromosphere emits increased levels of radiation during flares, which can be exceptionally energetic ($\sim$10$^{32}$ erg) events, initiated by a reconfiguration of the solar magnetic field. The magnetic reconnection which triggers flares typically originates in the Sun's tenuous corona, but result in energy being transported through the atmosphere towards deeper layers \citep{Benz2008, Fletcher2011}. The prevailing theory for flare energy propagation is by electrons accelerated along the reconnected field lines, which deposit their energy in the chromosphere via Coulomb collisions \citep{Korchak1967, Brown1971}. Additional energy transport and dissipation mechanisms, such as by high frequency Alfv\'{e}n waves have also been proposed \citep{Emslie1982, FletcherHudson2008, ReepRussell2016, Kerr2016}.

The majority of flare emission is generated in the lower atmosphere and is emitted in the visible and ultraviolet \citep{Kretzschmar2011, Fletcher2011, Milliganetal2014}. The extreme ultraviolet (EUV) region is particularly interesting during these events due to its variability, which can be as high as several orders of magnitude. This variability, observable in the Ly-$\alpha$ line, directly affects the Earth's atmosphere, resulting in detrimental effects on satellites and communication systems \citep{Woods2012, Kretzschmar2013}. It has also been established that flares drive flows in the chromosphere. These flows are typically believed to constitute a high-velocity upflow (``chromospheric evaporation") detectable in high-temperature species such as \ion{Fe}{19}, with an accompanying low-velocity downflow (``chromospheric condensation") in cooler species such as \ion{He}{2} and \ion{O}{5} \citep{Fisher1989, Milligan2006, Taroyan2014}. Observations of these flows are important in testing flare models as the speed, direction and duration of these flows are tied to the flare energy transport and deposition \citep{Fisheretal1985, Allred2005}.

\begin{figure*} 
	\figurenum{1}
	\bigskip
	\bigskip
	\plotone{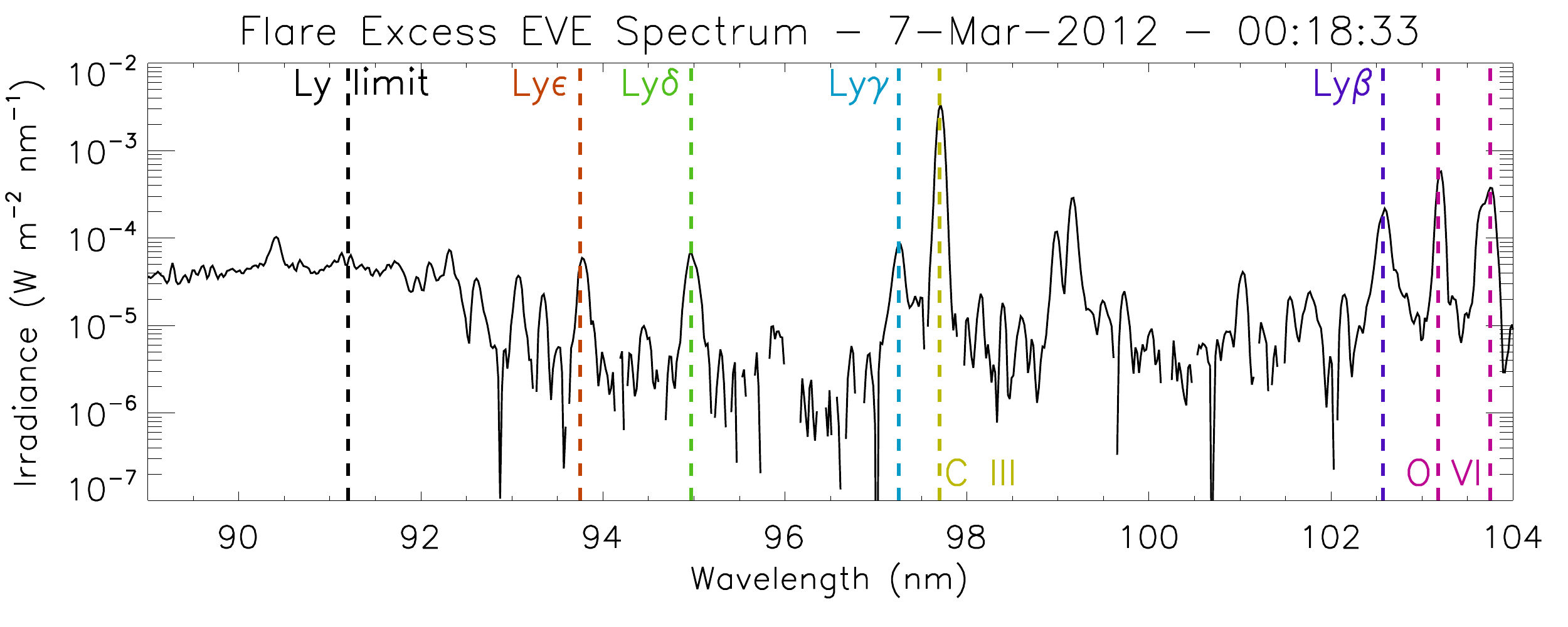}
	\caption{A MEGS-B spectrum of the hydrogen Lyman lines (after preflare subtraction) briefly after the onset of the X5.4 SOL2012-03-07T00:07 flare. The spectrum was obtained from Version 6 of the level 2 EVE data. Also prominent in this wavelength region are the \ion{C}{3} and \ion{O}{6} lines. \label{Figure1}}
\end{figure*}

The solar EUV output is monitored by the Extreme Ultraviolet Variability Experiment (EVE) instrument aboard the Solar Dynamics Observatory (SDO). EVE consists of two Multiple EUV Grating Spectrographs (MEGS); MEGS-A observes the wavelength range from 5 to 37 nm with a FWHM of around 0.1 nm, with an additional pinhole camera (MEGS-SAM) capable of measuring the region from 0.1 to 5 nm with a resolution of around 1 nm. MEGS-B observes the region from 35 to 105 nm, with an additional photodiode (MEGS-P) operating at 121.6 nm. As with MEGS-A, the MEGS-B detector has a resolution of roughly 0.1 nm, and a wavelength sampling of 0.02 nm. A MEGS-B flare spectrum is shown in Figure \ref{Figure1}. Unfortunately, MEGS-A is no longer operational as of May 2014, and MEGS-B functions on a reduced duty cycle of 3 hours per day. \citep{Woods2012, Brown2016}

Emission line spectroscopy of the hydrogen Lyman lines, which are well observed in the MEGS-B spectrum  ($ 91.2$ nm $< \lambda < 121.6$ nm), was performed by \citet{Brown2016}. In a sample of 6 M and X class flares, they found red- and blue- shifted Lyman lines suggesting plasma flows on the order of several tens of km s$^{-1}$, but with some flares showing upflows and some showing downflows. It could be the case that upflow signatures originate from eruptive features, but the observed Doppler speeds are much lower than the observed projected speeds of accompanying ejecta.

In this paper we explore the potential causes of red- and blue- shifted emission in the Lyman lines. We first use simulations computed by the RADYN code \citep{CarlssonStein1992, CarlssonStein1997, Allred2015} to simulate energy deposition into a model atmosphere by the injection of electron beams of varying properties. Two values for the beam spectral index ($\delta$) are used, allowing us to observe the response of a ``hard" beam ($\delta=3$) and a ``soft" beam ($\delta=8$). We then extract atmospheric snapshots from these simulations and use the RH code \citep{Uitenbroek2001} to calculate model line profiles, in which the important effects of partial redistribution (PRD) in the Lyman lines can be included in the radiative transfer. Both codes provide detailed predictions for the Lyman line profiles, which we then degrade by convolving with an approximation of the EVE instrumental profile. The comparison between these synthetic profiles and those observed by EVE reveals that the loss of detailed features, due to instrumental convolution, can lead to apparent Doppler shifts that mask the true flow direction of the Lyman lines.

\section{Numerical Tools} \label{sec:codes}

The RADYN and RH codes have been employed in many other studies, and are state-of-the-art resources to obtain model predictions for the shapes and intensities of spectral lines (and continua) emitted from solar and stellar flare atmospheres. Recently, \citet{Kuridze2015} probed the origin of the asymmetries in H$\alpha$ lines during flares. Synthetic H$\alpha$ and \ion{Ca}{2} 8542 \AA \ lines from RADYN were compared to IBIS observations of an M class flare by \citet{Rubiodacosta2015}. IBIS observations were again compared to synthetic \ion{Na}{1} D$_{1}$ profiles by \citet{Kuridze2016}. \citet{Simoes2016}  studied the formation of the He lines, and \citet{Kerr2016} investigated the formation of the \ion{Mg}{2} h \& k and \ion{Ca}{2} 8542 \AA \ lines. RADYN was also used to compare the ratio of H$\alpha$ to H$\beta$ line intensities in two simulations to IBIS observations of a C class flare, as reported by \citet{Capparelli2017}. Modelling of the spectra of M-dwarf flares has also been performed \citep{Allred2006, Kowalski2015, Kowalski2017}. Here, we outline some salient details of each code, but for full descriptions see \citet{Allred2015} and \citet{Uitenbroek2001}.

\subsection{RADYN}\label{subsec:radyn}

The RADYN code is a powerful and versatile tool for probing the response of an input atmosphere to the injection of energy. The code was developed by Carlsson \& Stein (\citeyear{CarlssonStein1992, CarlssonStein1997}) to study acoustic waves in the chromosphere, and was subsequently extended by \citet{AbbettHawley1999} to calculate the response to heating by a beam of non-thermal electrons (thus simulating a flare). Additional modifications to the code include a Fokker-Planck beam description, which more accurately models the diffusion of beam particles by pitch-angle scattering, and improved soft X-ray and extreme ultraviolet backwarming \citep{Allred2005, Allred2015}.

RADYN solves the non-linear, non-local equations of radiation hydrodynamics, that couple the hydrodynamic equations to the non-LTE radiative transfer equation and the non-equilibrium time dependent atomic level population equations. The non-LTE formalism is important in cases where the radiative rates contribute significantly to the level population density. Obtaining the correct level populations, and thus the correct emission and absorption coefficients, for the low-density chromosphere requires a non-LTE treatment.

The plane-parallel equations of radiative hydrodynamics are solved simultaneously, accounting for the conservation of mass, momentum, charge and internal energy density, along with the level population equation and the radiative transfer equation. These coupled equations are solved on a spatially-adaptive grid which dynamically adjusts to resolve strong gradients and shocks \citep{DorfiDrury}. Note that we used a modified version of RADYN with 300 grid cells rather than the typical 191. 

RADYN solves the level populations in our models for three elemental species that are of paramount importance in describing the radiation field in the chromosphere and transition region. These consist of a six-level plus continuum hydrogen atom, a nine-level plus continuum helium atom, and a six-level plus continuum \ion{Ca}{2} ion. Inclusion of the continuum allows for the calculation of bound-free transitions in addition to the bound-bound transitions. Transitions are computed for 5 different viewing angles with up to 201 frequency points, and are done so with the assumption of complete redistribution (CRD) which assumes that a photon undergoing a scattering or absorption is re-emitted with a wavelength that is uncorrelated to its original wavelength. In RADYN, we mitigate the effects of CRD by modelling the Lyman lines as Doppler profiles \citep{Leenaarts2012}, but a more accurate approach would be to use partial redistribution (PRD).

\subsection{RH}\label{subsec:rh}

The radiative transfer code RH, developed by \citet{Uitenbroek2001}, allows the computation of spectral line profiles with the inclusion of PRD effects. Based on the accelerated lambda iteration (ALI) method for multilevel atoms \citep{RybickiHummer1991}, it solves the equations of statistical equilibrium and radiative transfer.

While the treatment of chromospheric lines with the assumption of CRD is computationally advantageous for RADYN, it has been demonstrated this formalism is not accurate for the Lyman lines \citep{Vernazza1973, HubenyMihalas2014}. As mentioned above, CRD assumes lack of coherence between the absorbed and emitted photon (due to collisions). However, in low-density environments like parts of the solar chromosphere there may be insufficient collisions before a photon is re-emitted. In these cases, the coherence is not destroyed, and the wavelength of the emitted photon is correlated with that of the absorbed photon. Assuming CRD may lead to overestimation of the line wing intensity, as a photon absorbed in the line core can be re-emitted in the wing, whereas in PRD a photon absorbed in the core will more likely be re-emitted with a wavelength close to the core. CRD may also lead to an inaccurate number of wing photons being scattered into the core, whereas in PRD they will more readily escape as the wing is more optically thin.

\citet{Vernazza1973} could not replicate observations of the quiet-Sun Ly-$\alpha$ line without computing it with a mixture of CRD and PRD. \citet{Rubiodacosta2015} also note that the application of PRD is desirable not only for correct computation of the Lyman lines, but also for H$\alpha$ as this line shares an upper level with Ly-$\beta$. Similarly, \citet{Uitenbroek2002} found that the CRD approximation led to overestimations of the radiative rates in the solar \ion{Ca}{2} K line. For these reasons, we use RH to obtain Lyman line profiles under the assumption of PRD.

Since RH is a time-independent code, we input atmospheric snapshots obtained from RADYN simulations to the 1D version of RH, building a dynamic picture from these individual times. We note that in this approach we are in effect neglecting the ``history" of the atmosphere since RH re-solves the hydrogen populations in statistical equilibrium. We input the non-equilibrium electron density as computed from RADYN, which mitigates this to some extent. Additionally, non-thermal collisions with the beam are neglected. The non-LTE calculations are performed for any atomic species included as ``active", and an LTE assumption is used for species included as background (``passive").

\section{Description of simulations and methods} \label{sec:runs}

Simulations from the RADYN code were used to emulate a variety of flare types, which are differentiated by the parameters of the electron beam. The beam is characterised by its low-energy cutoff (E$_{c}$), spectral index ($\delta$), and non-thermal energy flux. The flux varies in time, with a 20s triangular profile, peaking at t=$10$ s in all simulations. The $\delta$ and E$_{c}$ values effectively determine the altitude of the energy deposition. A high $\delta$ means that the number of beam electrons drops off sharply as a function of energy, therefore the distribution is weighted to the lower-energy electrons which are stopped higher up in the atmosphere. Conversely, a low $\delta$ beam contains more high-energy electrons which can penetrate and heat deeper into the chromosphere.

\begin{deluxetable}{c c c c c}
	\tablenum{1}
	\tablecolumns{4}
	\tablecaption{List of RADYN simulationss used \label{Table1}}
	\tablehead{
		\colhead{Beam}\tablenotemark{a} & \colhead{F$_{peak}$ (erg cm$^{-2}$ s$^{-1}$)} & \colhead{$\delta$} & \colhead{E$_{c}$ (keV)} & \colhead{F$_{tot}$ (erg cm$^{-2}$)}}
	\startdata
	F10D3 & 1x10$^{10}$ & 3 &  25 & 1x10$^{11}$  \\
	F10D8 & 1x10$^{10}$ & 8 &  25 & 1x10$^{11}$  \\
	3F10D8 & 3x10$^{10}$ & 8 & 25 & 3x10$^{11}$ \\
	F11D3 &  1x10$^{11}$ & 3 &  25 & 1x10$^{12}$  \\
	\enddata
	\tablenotetext{a}{The simulations used are publicly available on the grid of RADYN models at \url{https://star.pst.qub.ac.uk/wiki/doku.php/public/solarmodels/start}. The F10D3, F10D8, 3F10D8 and F11D3 simulations have model numbers 55, 60, 66 and 67 respectively.}
\end{deluxetable} 

We focus on four particular simulations from RADYN: two that deposit energy rather deep into the chromosphere with moderate and high beam fluxes, and two that deposit a greater fraction of their energy at a higher altitude in the atmosphere. All simulations were obtained from the online grid of RADYN models and are freely available courtesy of the European Commission funded F-CHROMA collaboration (\url{https://star.pst.qub.ac.uk/wiki/doku.php/public/solarmodels/start}). The properties of each simulation are described in Table \ref{Table1}. Each electron beam was injected into a loop of half-length $10$ Mm.

The intial pre-flare atmosphere in each of the simulations was a VAL3C-like atmosphere \citep{Vernazza1981}. Our starting atmosphere was constructed from the VAL3C temperature structure, from which the heating required to sustain that temperature structure was computed. This atmosphere plus heating function was then allowed to relax to an equilibrium state, so that the upper chromosphere differs somewhat from VAL3C (M. Carlsson, private communication).

We then explore line formation by following the approach of \citet{CarlssonStein1997}, examining the contribution function to the emergent intensity ($C_{I}$), which is given by the integrand in Equation 1. Integrating $C_{I}$ over height in the atmosphere results in the emergent intensity ($I_{\nu}$), meaning that $C_{I}$ effectively tells us the locations that contribute to line formation in the atmosphere. We use $C_{I}$ to describe features observed in each simulation at key times further on in this paper. 
 
 \begin{equation}
 I_{\nu}=\int_{z_{0}}^{z_{1}} S_{\nu} \tau_{\nu}e^{-\tau_{\nu}} \frac{\chi_{\nu}}{\tau_{\nu}} dz
 \end{equation}

 Due to the assumption of CRD in RADYN, $S_{\nu}$ is constant across a line profile, but does vary as a function of height. In RH we employ PRD and so $S_{\nu}$ is a function of frequency. The $\tau_{\nu}e^{-\tau_{\nu}}$ component describes how radiation is attenuated as a function of height for a given frequency, and is large when $\tau=1$. The final component, $\frac{\chi_{\nu}}{\tau_{\nu}}$, is large when there are many emitting particles at a low optical depth (where $\chi_{\nu}$ is the monochromatic opacity per unit volume), which highlights flows in the atmosphere \citep{CarlssonStein1997}.

The RADYN outputs are then prepared for input to RH. This is done by decomposing the time-resolved atmospheric arrays into multiple atmospheric ``snapshots" which span the duration of each simulation. These snapshots define the temperature ($T$), electron density ($n_{e}$), macroscopic velocity ($V_{z}$) and microturbulent parameter ($V_{turb}$=2 km s$^{-1}$) on a column mass depth scale. Note that we input the non-equilibrium electron density into RH, which somewhat mitigates using statistical equilibrium to obtain the hydrogen level populations. RH is then run with each of these atmospheres in sequence, using a 6-level hydrogen atom with a continuum level. Each of the Lyman lines are treated with PRD effects.

The final step in analysing the output line profiles from RADYN and RH is to simulate instrumental effects and measure the Doppler shifts that an instrument such as EVE would observe \citep{Brown2016}. This is done by first rebinning the non-constant RH data (between 30 and 130 nm) to a constant wavelength spacing of 0.005 nm, before convolving the RADYN \& RH profiles with the instrumental profile of EVE, given by a Gaussian of FWHM 0.085 nm (see \citet{Crotser2007}) and then resampling the data to replicate EVE's wavelength bin size of 0.02 nm. Measurements of the line centroid variations are performed by both Gaussian fitting and intensity weighted means. This is done for both the RADYN output and the RH output, providing two sets of velocity results for each simulation.

\section{Results} \label{sec:results}
\subsection{The F10D3 simulation}\label{subsec:f10results}

\begin{figure*} 
	\figurenum{2}
	\plotone{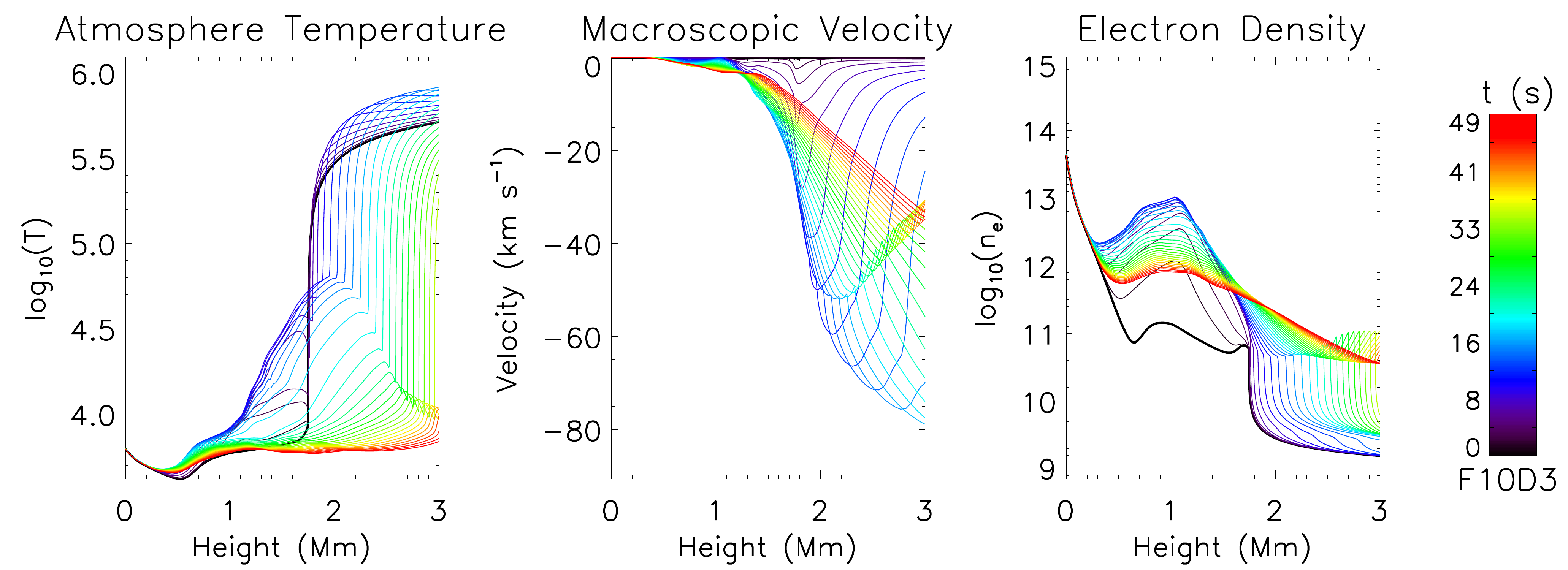}
	\caption{The atmospheric variables for the F10D3 simulation, with the pre-flare atmosphere plotted in black. The evolution of the atmosphere is represented by the varying line colours. Negative velocities correspond to upflows. Quantities are plotted at $1.5$ s intervals. \label{Figure2}}
\end{figure*}

The F10D3 simulation ($\delta=3$, E$_{c}=25$ keV) describes a moderate amount of heating spanning the first 20 seconds, followed by 30 seconds of cooling and relaxation. The evolution of the atmosphere is shown in Figure \ref{Figure2}, which shows that the atmospheric temperature is quick to respond to the beam injection and increases at all heights around the chromosphere and transition region (which rises to a higher altitude). Over the duration of the flare the atmosphere expands, with the transition region settling at an altitude $> 3$ Mm, by $t=50$ s. The chromosphere now extends over a greater height range than the pre-flare atmosphere. $n_{e}$ remains elevated by almost two orders of magnitude between z=2-3 Mm above the pre-flare value, which results in significant cooling via a high amount of radiative losses.

The temperature increase at early times is accompanied by an atmospheric upflow that attains a speed of 80 km s$^{-1}$ at $z=3$ Mm. The beam injection also causes a net increase in the overall electron density through a combination of the rise in temperature, and a significant number of non-thermal collisions with the beam itself.

\begin{figure*} 
	\figurenum{3}
    \gridline{\fig{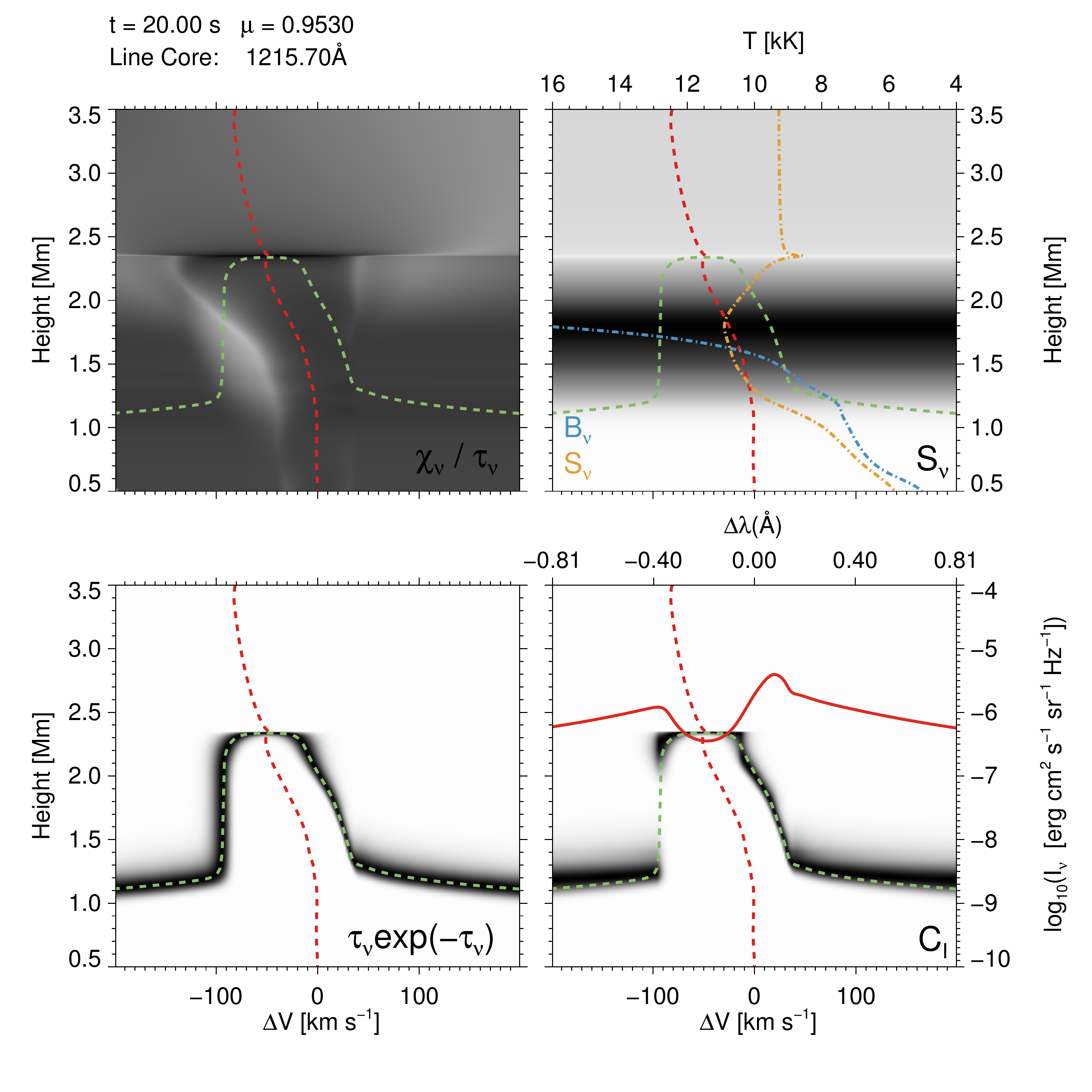}{0.45\textwidth}{(a)}
             	 \fig{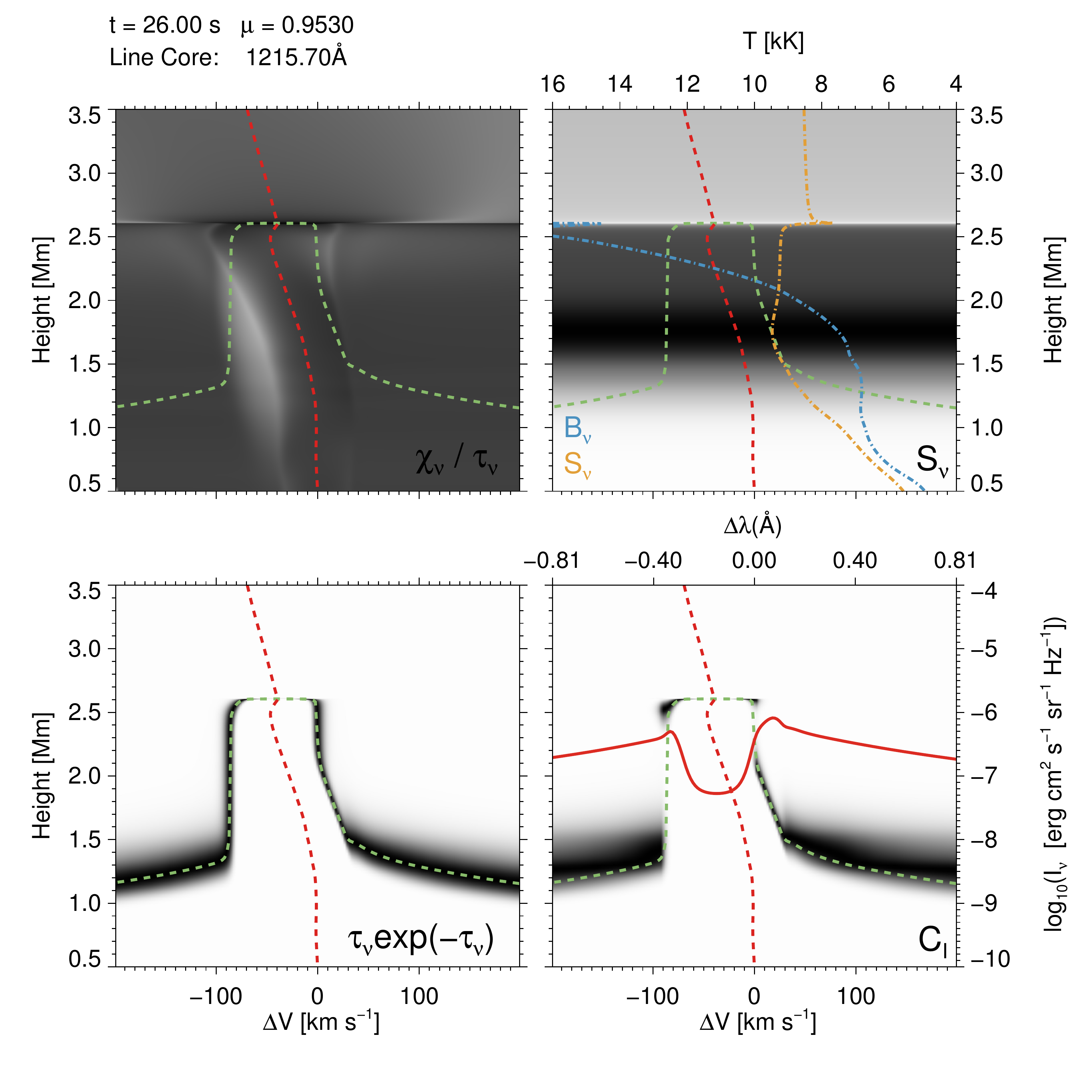}{0.45\textwidth}{(b)}}
    \gridline{\fig{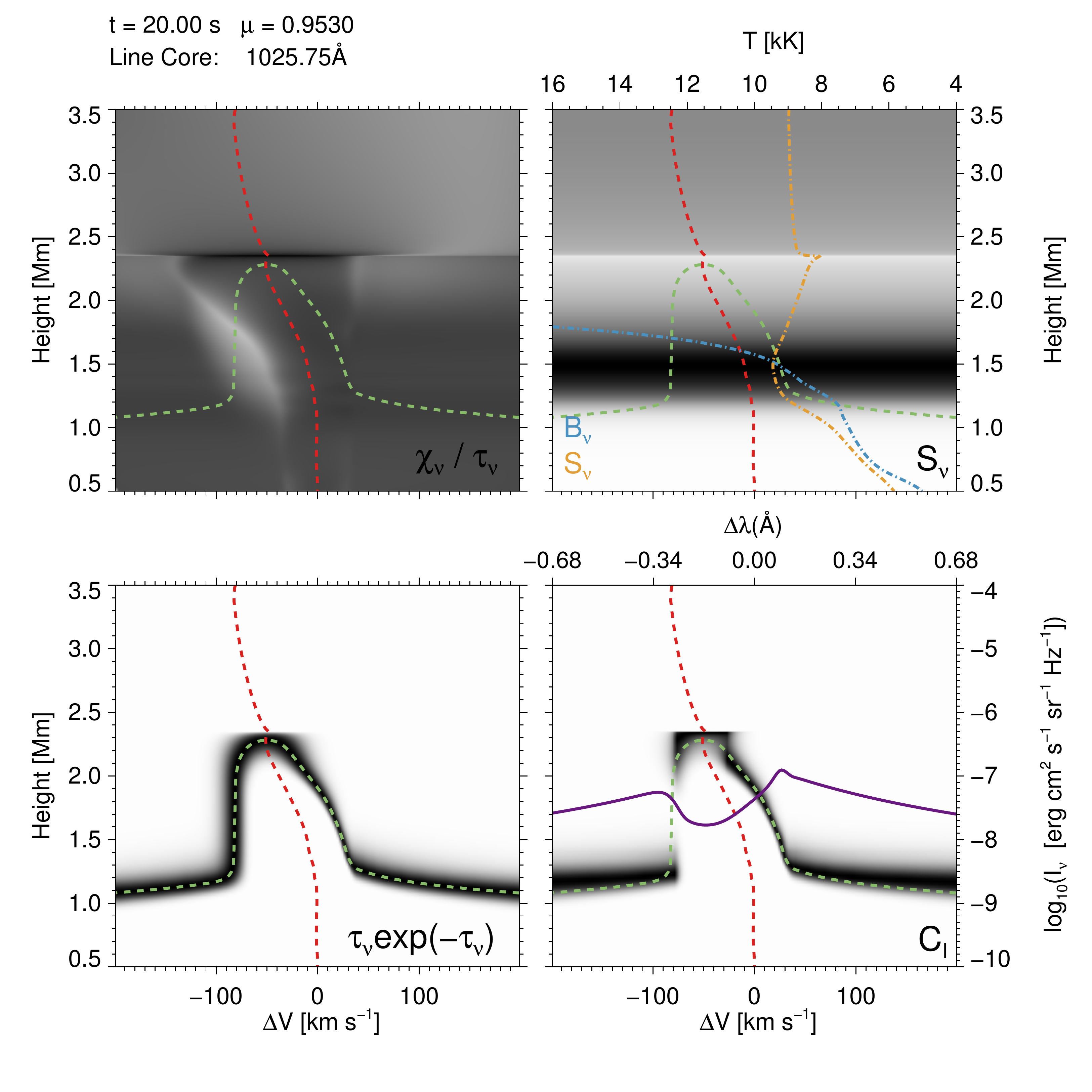}{0.45\textwidth}{(c)}
              	\fig{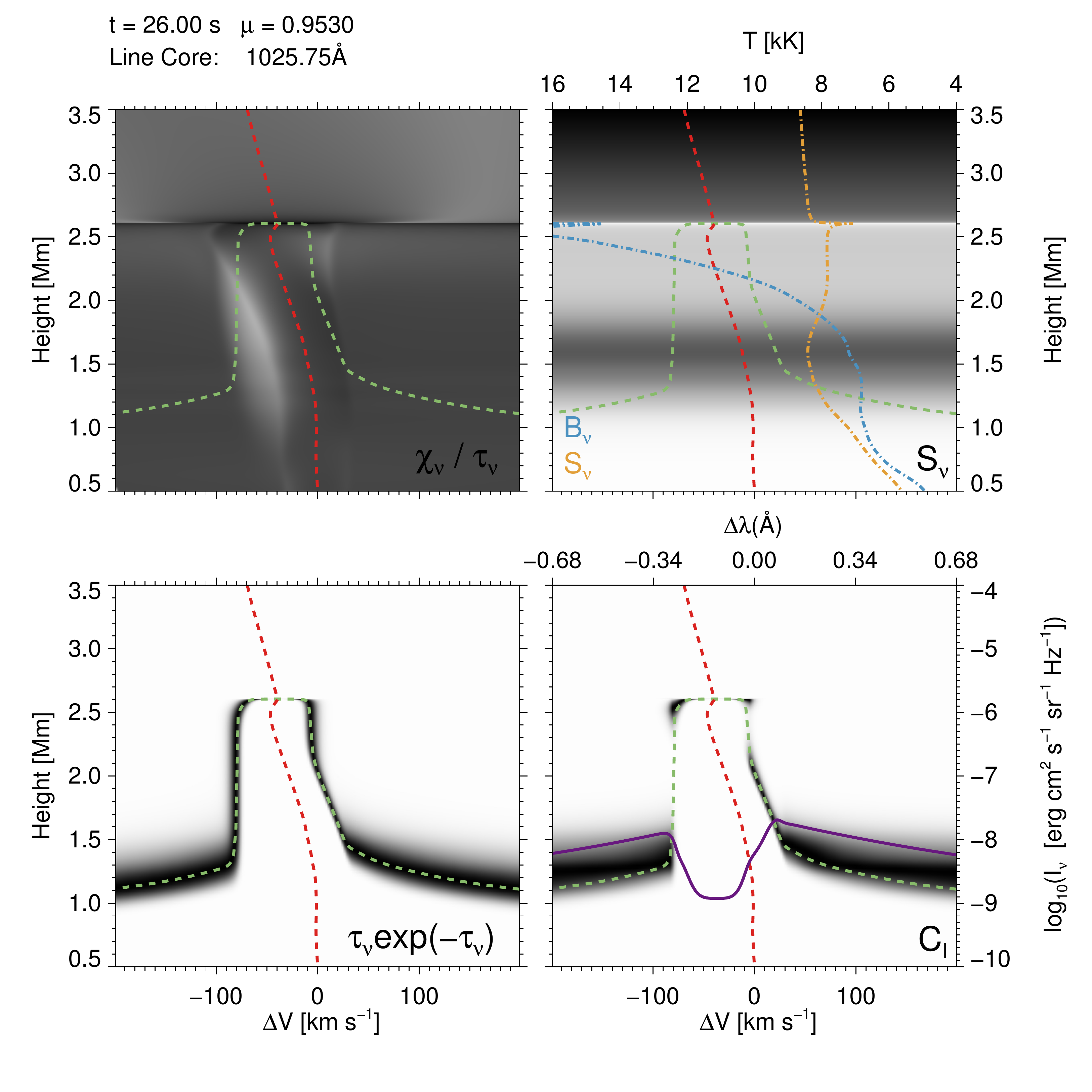}{0.45\textwidth}{(d)}}         	
       	\caption{Line contribution functions for Ly-$\alpha$ (a \& b) and Ly-$\beta$ (c \& d) at two times during the F10D3 simulation corresponding to the end of the energy deposition (t = 20 s) and atmospheric relaxation (t = 26.0 s). Each subfigure images a separate constituent of the overall contribution function, which is stated in the lower right of each panel. Darker colors indicate higher values of their respective quantities. The dashed red and green lines show the atmospheric velocity and $\tau_{\nu}$ = 1 surface respectively. The dot-dashed blue and yellow lines respectively map the Planck and source functions as a function of height, in units of radiation temperature. The emergent line intensities are overplotted on the lower right panels in red and purple.\label{Figure3}}
\end{figure*}

We present the line contribution functions from RADYN for Ly-$\alpha$ and Ly-$\beta$ in Figure \ref{Figure3}. At $t=20$ s the electron beam has just stopped heating the atmosphere. At this time, Figures \ref{Figure3}a and \ref{Figure3}c show that both Ly-$\alpha$ and Ly-$\beta$ have developed a blue asymmetry in their $\tau_{\nu}=1$ surfaces, peaking blueward of the rest wavelength at a height around 2.3 Mm. The lower right hand panels of these figures indicate that the line contribution function forms around the $\tau_{\nu} = 1$ height, confirming that the lines are optically thick during the flare. Here we introduce the definition that the core of the line is the part of the line that forms the highest in the atmosphere (similar to \citet{Rathore2015}). The line wings form lower in the atmosphere because wing photons are not as readily absorbed, and can escape more easily.

 Since the line core forms in a region of upflowing plasma (with velocities of 40-50 km s$^{-1}$), the opacity structure is also shifted to the blue and the emergent profiles exhibit a blueshifted core.

 The line profiles appear similar to each other, with both Ly-$\alpha$ and Ly-$\beta$ exhibiting central reversals. The central reversals are a consequence of the line source functions having local maxima closer to the wing formation height (around $z=1.5-1.8$ Mm) as opposed to the core formation height, resulting in stronger emission at the wing frequencies as opposed to the line core, which forms at an altitude where $S_{\nu}$ is relatively weaker.

The blueshift in the $\tau_{\nu}$=1 surface causes the central reversal to also be shifted to the blue, and a distinct asymmetry in the line's central reversal can be seen in both Ly-$\alpha$ and Ly-$\beta$ which effectively acts to reduce the amount of emission in the blue wing relative to red wing.

By $t=26$ s the transition region has moved upwards through the atmosphere to 2.6 Mm, and the $\tau_{\nu}=1$  surfaces for both Ly-$\alpha$ and Ly-$\beta$ show that the line cores are now formed at the top of the chromosphere, with the line formation region now spanning a greater range in height. The atmosphere is still upflowing at the core-formation height, and consequently the $\tau_{\nu}$ = 1 surface maintains a blue asymmetry while the line core is still blueshifted. $S_{\nu}$ has decreased in the region encompassing the line formation, thus the line profiles are weaker in intensity. They still retain a central reversal, due to $S_{\nu}$ in the line formation region being weaker at the core formation height than at lower altitudes. As a result of the asymmetric $\tau_{\nu}$ = 1 surface, the central reversals are still strongly shifted to the blue.

\begin{figure*} 
	\figurenum{4}
	\plotone{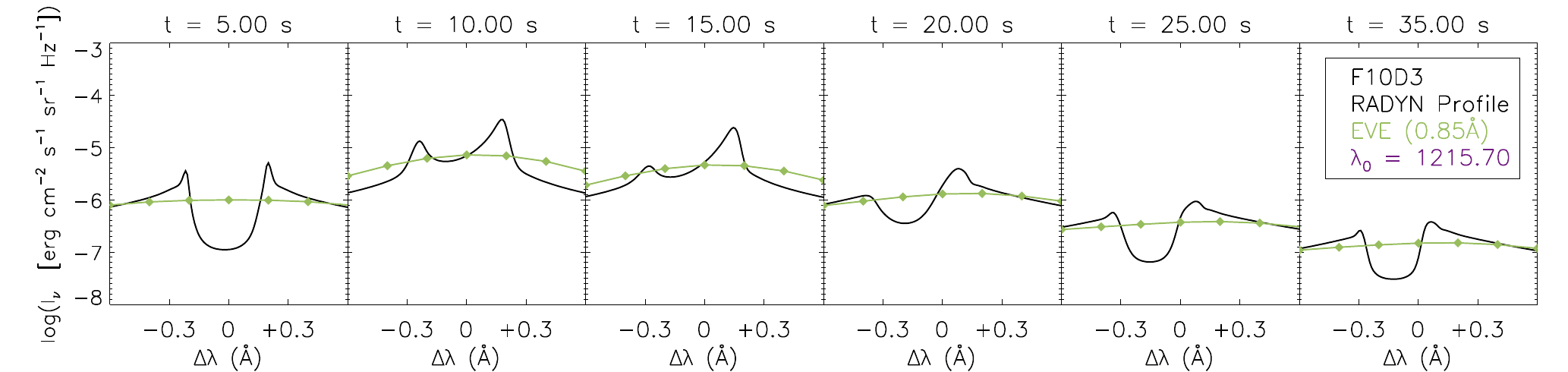}
	\caption{The Ly-$\alpha$ line at various times during the F10D3 simulation before (in black) and after (in green) instrumental convolution. This involves smoothing of the line by a Gaussian of width 0.85 {\AA}  before a rebinning to a wavelength sampling of 0.2 \AA. \label{Figure4}}
\end{figure*}

To understand how the detailed line profiles from the F10D3 simulation would be recorded by the EVE instrument, we perform the rebinning and Gaussian convolution procedures described in Section \ref{sec:runs}. to emulate the EVE instrumental profile \citep{Crotser2007} and compare the simulated velocities as they would be observed by EVE to those reported in \citet{Brown2016}. Examples of degraded line profiles for Ly-$\alpha$ are shown in Figure \ref{Figure4}, and clearly display a stark reduction in detail compared to the raw data.

 \begin{figure*} 
 	\figurenum{5}
 	\plotone{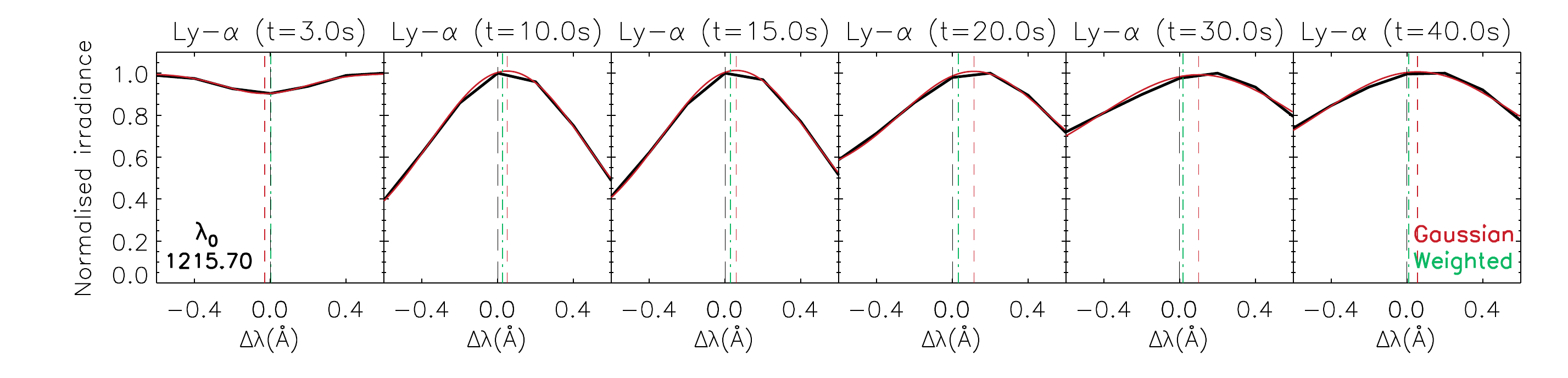}
 	\caption{Gaussian fits (in red) to the degraded Ly-$\alpha$ line profiles (in black) as in Figure \ref{Figure4} at various times throughout the F10D3 simulation. Line centroid positions derived from the Gaussian fit are indicated by the dashed red lines, and those derived from intensity weighting by the broken green lines.\label{Figure5}}
 \end{figure*}

\begin{figure*} 

	\figurenum{6}
	\gridline{\fig{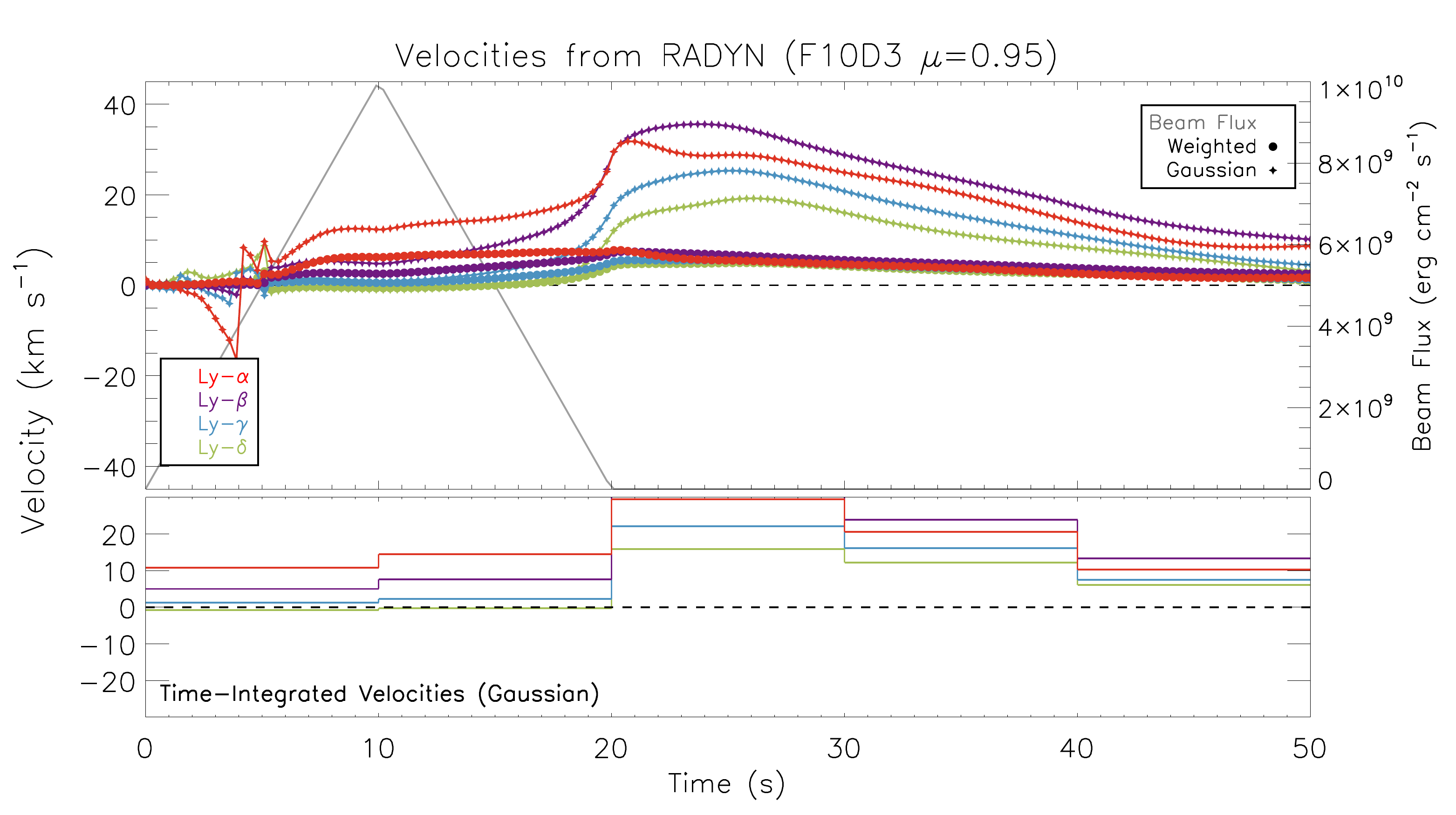}{0.56\textwidth}{(a)}
		\fig{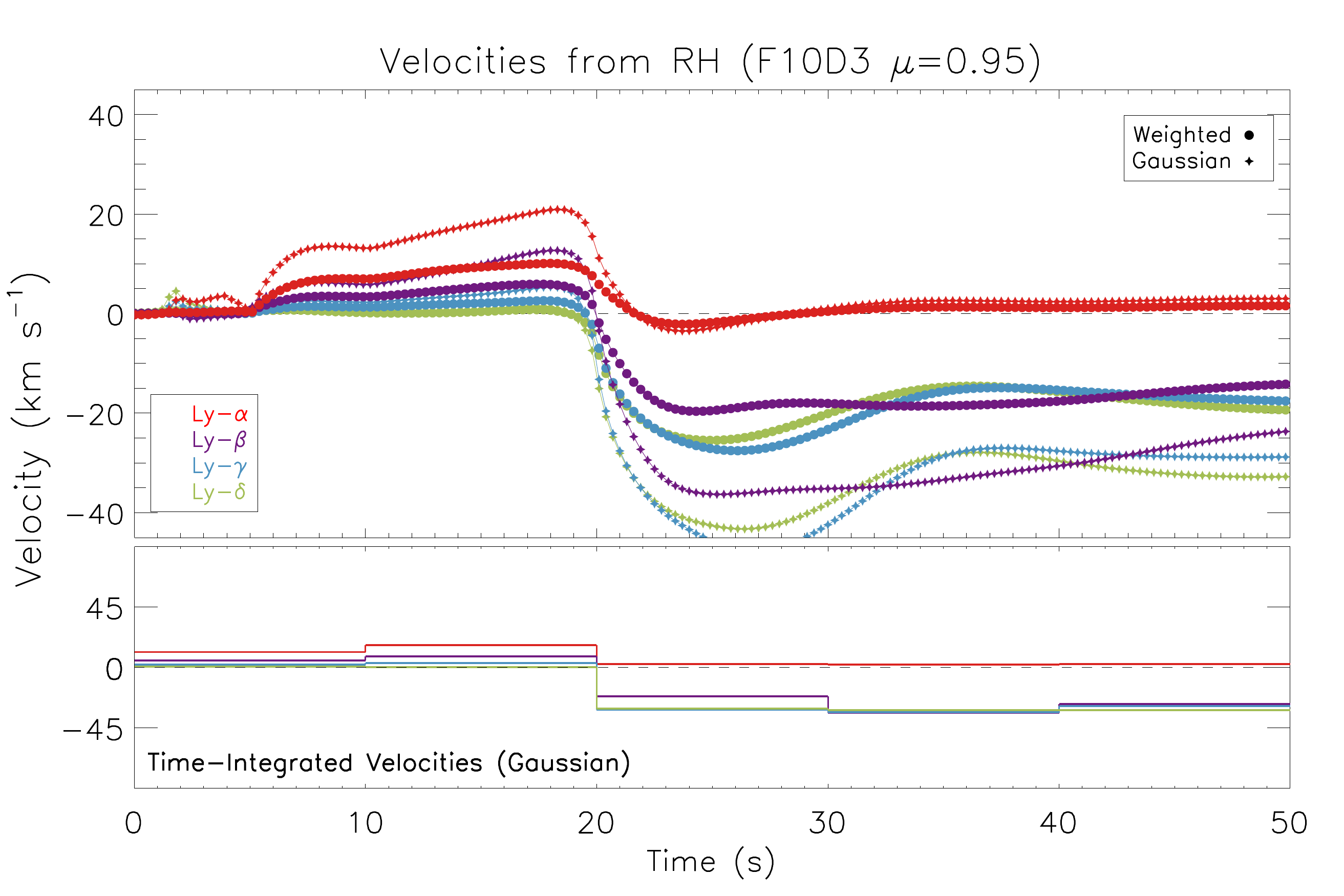}{0.47\textwidth}{(b)}}
	\caption{Doppler velocities of the F10D3 line profiles obtained by emulating the instrumental effects and simulating observations from the EVE instrument. Velocities are displayed for the RADYN (a) and the RH (b) profiles, with positive velocities indicating downflows. Circular data-points were obtained from the intensity-weighting method, and those plotted with a diamond symbol were obtained from Gaussian fitting. The lower panels also show Doppler velocities, but with the line irradiances time-integrated for 10 s before Gaussian fitting is performed in order to fully emulate an EVE observation. \label{Figure6}}
\end{figure*}

\subsubsection{Velocities from RADYN}

Velocities are obtained by measuring the deviation of the line centroid positions from their rest wavelengths. These measurements are achieved by two methods; fitting a 4-parameter Gaussian to the line profile and obtaining the measured line centroid, and by averaging the wavelength values weighted by the intensity at each wavelength bin. These methods are visualised in Figure \ref{Figure5}.  Velocity profiles for Ly-$\alpha$ through Ly-$\delta$ are plotted as a function of time in Figure \ref{Figure6}a. We also include a panel of velocities calculated on the time-integrated line profiles (for 10 s) in order to facilitate a comparison to EVE observations \citep{Brown2016}. The time-integrated velocities demonstrate an additional loss of information due to the instrumentation when compared to the results with high temporal resolution.

Each of the lines plotted in Figure \ref{Figure6}a exhibits redshifts of varying magnitude at $t=20$ s, with Ly-$\alpha$ and Ly-$\beta$ displaying the most prominent shifts during the beam deposition. Once the beam stops heating the atmosphere (after $t=20$ s), the redshift signatures quickly peak, suggesting downflows of $30-35$ km s$^{-1}$ in the Ly-$\alpha$ and Ly-$\beta$ lines when the Gaussian fitting method is used. Ly-$\gamma$ and Ly-$\delta$ exhibit redshifts corresponding to flows of $20-25$ km s$^{-1}$. There is a general ordering to the derived velocities throughout the simulation, with lower order lines suggesting higher speeds. This is particularly interesting as the opposite effect was observed in velocity profiles with no preflare-subtraction by \citet{Brown2016} when ordering was present, although this could have been due to the variability in the measured velocities of the higher-order EVE lines also being greater due to their weaker irradiances.

It is illuminating that the velocity signatures from the degraded profiles suggest downflows throughout the duration of this simulation. From Figures \ref{Figure3}a and \ref{Figure3}c, it is clear that the beam-heating stage results in blueshifts in the line cores. However, because the line profiles are centrally reversed and shifted to the blue, the red peak becomes dominant. The red peak drags the derived line centroids further red than the line core. Similarly, during relaxation, it can be seen from Figures \ref{Figure3}b and \ref{Figure3}d that the persistence of the blueshift in the line core maintains an overall red asymmetry in each of the lines, meaning that the velocity profiles continue to exhibit redshifts until the end of the simulation.

It should also be noted that the Doppler shifts observed between t=0-5 s are an artefact of the instrumentally-convolved line profiles transitioning from absorption profiles to emission profiles, which skews the Gaussian fit as the line profiles briefly flatten. The initial absorption profiles do not persist for as long in the higher flux simulations. 

The cause of the redshifts observed in Figure \ref{Figure6}a is less obvious once the profiles undergo degradation by the instrument, and it would be easy for the line profiles to be misinterpreted as emitting an excess in the red wing as opposed to being highly absorbing in the blue wing.

\subsubsection{Velocities from RH}\label{subsubsec:rhvelf10}

The atmosphere snapshots from the RADYN simulation were also used as input to RH. These were run with a 6 level-with-continuum hydrogen atom with each of the Lyman lines computed with PRD. The output RH profiles were then degraded as before and the simulated Doppler velocities were calculated. These velocities are displayed in Figure \ref{Figure6}b. 

The RH velocity profiles agree rather well with those obtained from RADYN throughout the beam-heating stage ($t=0-20$ s), with redshifts found in each of the lines and Ly-$\alpha$ and Ly-$\beta$ again displaying the more prominent signatures. At $t=20 s$, Ly-$\alpha$ suggests a peak downflow speed of $20$ km s$^{-1}$. Significant differences arise between RADYN and RH after the electron beam is switched off. In RH, the shift in Ly-$\alpha$ decays to zero, while the other Lyman lines abruptly transition into exhibiting strongly blueshifted signals. At $t=26$ s, the blueshifted signatures peak, with Ly-$\gamma$ and Ly-$\delta$ now exhibiting the strongest flows with the former reaching speeds of around $50$ km s$^{-1}$. These signatures then decay over the remainder of the simulation.

\begin{figure*} 
	\figurenum{7}
	\plotone{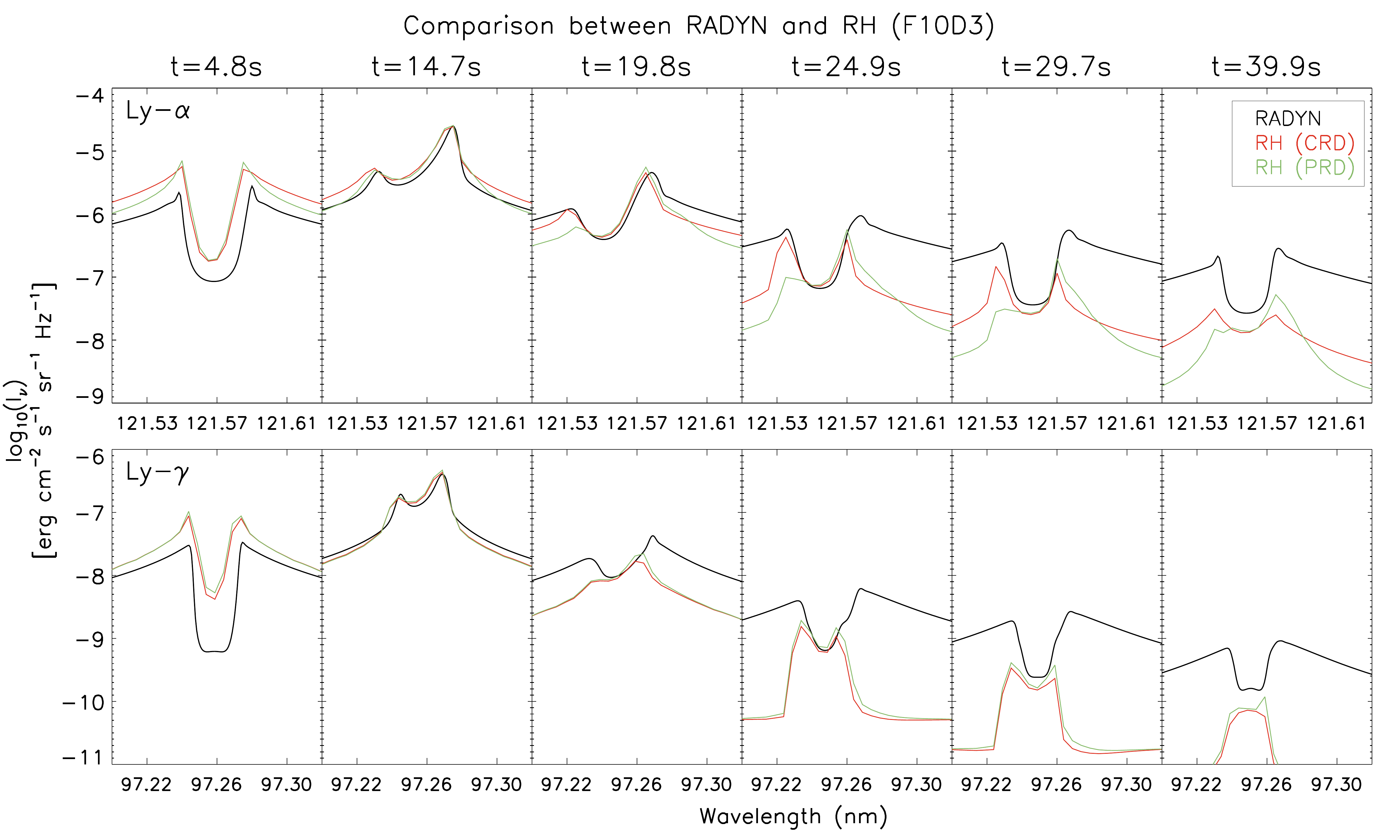}
	\caption{Ly-$\alpha$ and Ly-$\gamma$ line profiles obtained from the F10D3 RADYN simulation, with RH output calculated with both CRD and PRD formalisms as a function of time. To help illustrate the features causing the differences between Figure \ref{Figure6}a and b, we do not show these profiles with instrumental effects.  \label{Figure7}} 
\end{figure*}

To understand why the velocity profiles deviate so significantly from each other after t=20 s, a comparison between the RADYN and RH profiles for Ly-$\alpha$ and Ly-$\gamma$ is shown in Figure \ref{Figure7}. We include Ly-$\gamma$ as it exhibits the strongest blueshifted signatures in Figure \ref{Figure6}b. In addition to the RH profiles computed with PRD, we include those obtained from RH when CRD is applied. It can be seen that after t=20 s, both Ly-$\alpha$ and Ly-$\gamma$ as computed from RH have much weaker wing intensities than the RADYN profiles, while the core intensities show better agreement. 
	
The drop in the wing intensities in Ly-$\alpha$ from RH is not as pronounced as in Ly-$\gamma$. The consequence of this is that the RADYN profiles remain centrally-reversed as a result of the high wing intensities, while the RH profiles, although still centrally-reversed, are more prominently peaked due to the lower intensities in the far wings (moreso in the higher order lines). This means that while RADYN continues to produce redshifted profiles as a result of blueshifts acting in the centrally-reversed line cores, the blueshifts in the RH profiles for Ly-$\beta$ through Ly-$\delta$ act on more emissive features and therefore strengthen the blue wing, meaning that upflows are obtained in the velocity profiles.

While the application of PRD was expected to produce line profiles that differed from those computed by RADYN, it is evident from Figure \ref{Figure7} that even the RH profiles computed using CRD can deviate from their RADYN counterparts, with significant differences between the RADYN and RH (CRD) profiles after the beam heating stops. Further, the CRD and PRD results computed by RH were surprisingly similar during the main phase of the flare, suggesting an alternate cause for the difference between synthetic spectra from RH and RADYN.

RH computes the atomic level populations using the equations of statistical equilibrium, whereas RADYN employs non-equilibrium excitation and ionisation, both of which will affect the resulting line and continuum emission, with the latter being important for emission far from the line core. It is therefore possible that the majority of the differences between the RADYN and RH profiles arise from the alternative methods used for obtaining the level populations, and not because of the application of PRD.

 While the beam is being injected, the additional heating and direct excitation of the plasma by the beam acts to increase the amount of collisional excitation, which enhances the populations of the upper levels.  This also increases the recombination rate as $n_{e}$ increases. As a result, the conditions in RADYN closely approximate statistical equilibrium.  However, once the beam switches off, the dynamics of the atmosphere continue to evolve rapidly, while the level populations may take tens of seconds to return to equilibrium levels \citep{CarlssonStein2002}. This means that statistical equilibrium becomes a poor approximation once the beam is switched off, leading to differences in the computed RH profiles with respect to RADYN. Further investigation is required, but it iseems likely that the enhanced electron density facilitates more collisions, so that the PRD solution approaches the CRD solution.

Across both sets of results in Figure \ref{Figure6}, the maximum speeds of the apparent flows lie roughly between $20-40$ km s$^{-1}$. This is in line with previous velocity measurements for chromospheric lines \citep{MilliganDennis2009, Brown2016}. However, the apparent flow direction does not reflect that of the atmospheric velocity because the line profiles are centrally reversed. From Figure \ref{Figure3}, the core formation height is upflowing with a velocity of $50$ km s$^{-1}$ at $t=20$ s, while Figure \ref{Figure6} shows downflows at this time. Velocities obtained from intensity-weighting are not as large as those obtained from Gaussian fitting, but do verify the direction of the observed shifts.

\subsection{The F10D8 simulation}\label{subsec:f10d8results}

To assess the impact of injecting energy at higher altitude, we increased the spectral index to a value of 8 while keeping other parameters fixed. Consequently, this electron beam has a sharper drop off of electron number as a function of energy, and therefore is constituted of a greater number of lower-energy electrons which are stopped at higher altitudes in the model atmosphere.

\begin{figure*} 
	\figurenum{8}
	\plotone{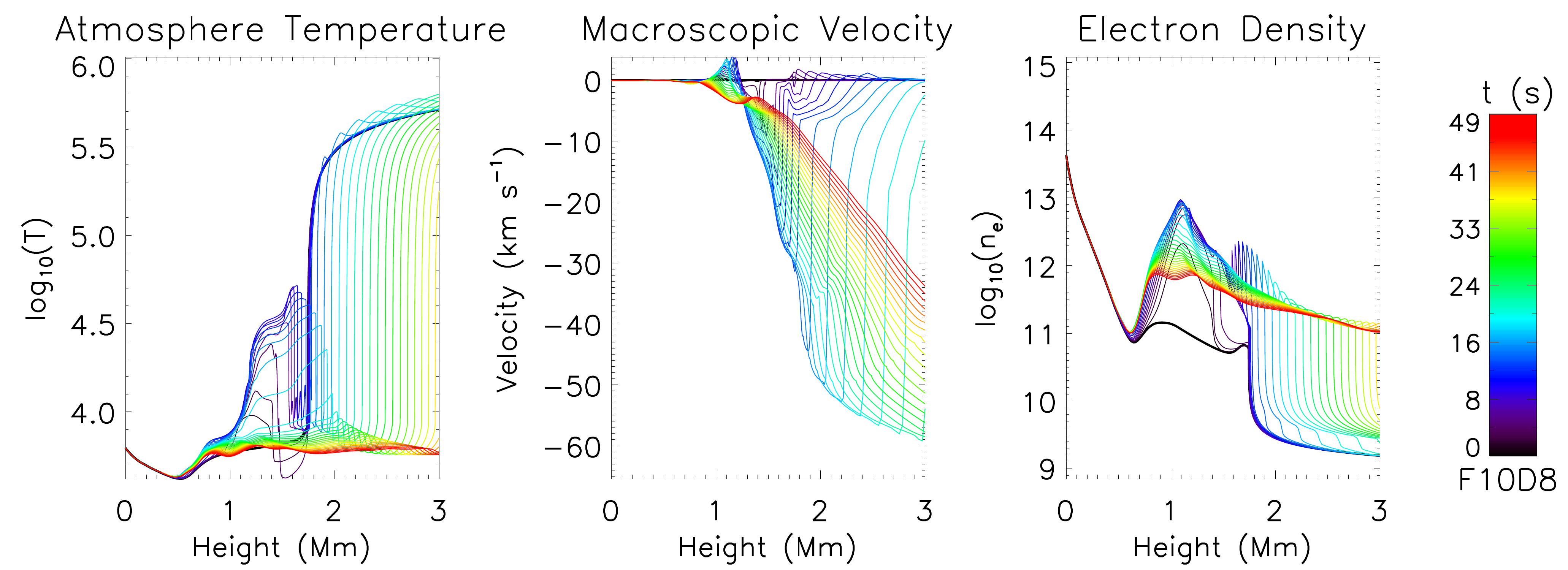}
	\caption{Atmospheric evolution during the F10D8 simulation. The pre-flare atmosphere is indicated by the thick black line. Quantities are plotted at $1.5$ s intervals. \label{Figure8}}
\end{figure*}

The evolution of the atmosphere in this simulation is shown in Figure \ref{Figure8}. Compared to the F10D3 simulation (Figure \ref{Figure2}), the lower atmosphere (below z=1.7 Mm) develops a much steeper temperature gradient throughout the first 20 s, which dissipates as the transition region moves upwards. The steep temperature gradients are cospatial with localised enhancements in the electron density. Upflows are again initiated by the beam injection, with velocities in the lower chromosphere slightly lower (around 60 km s$^{-1}$) than that of the F10D3 simulation.

\begin{figure*} 
	\figurenum{9}
	\gridline{\fig{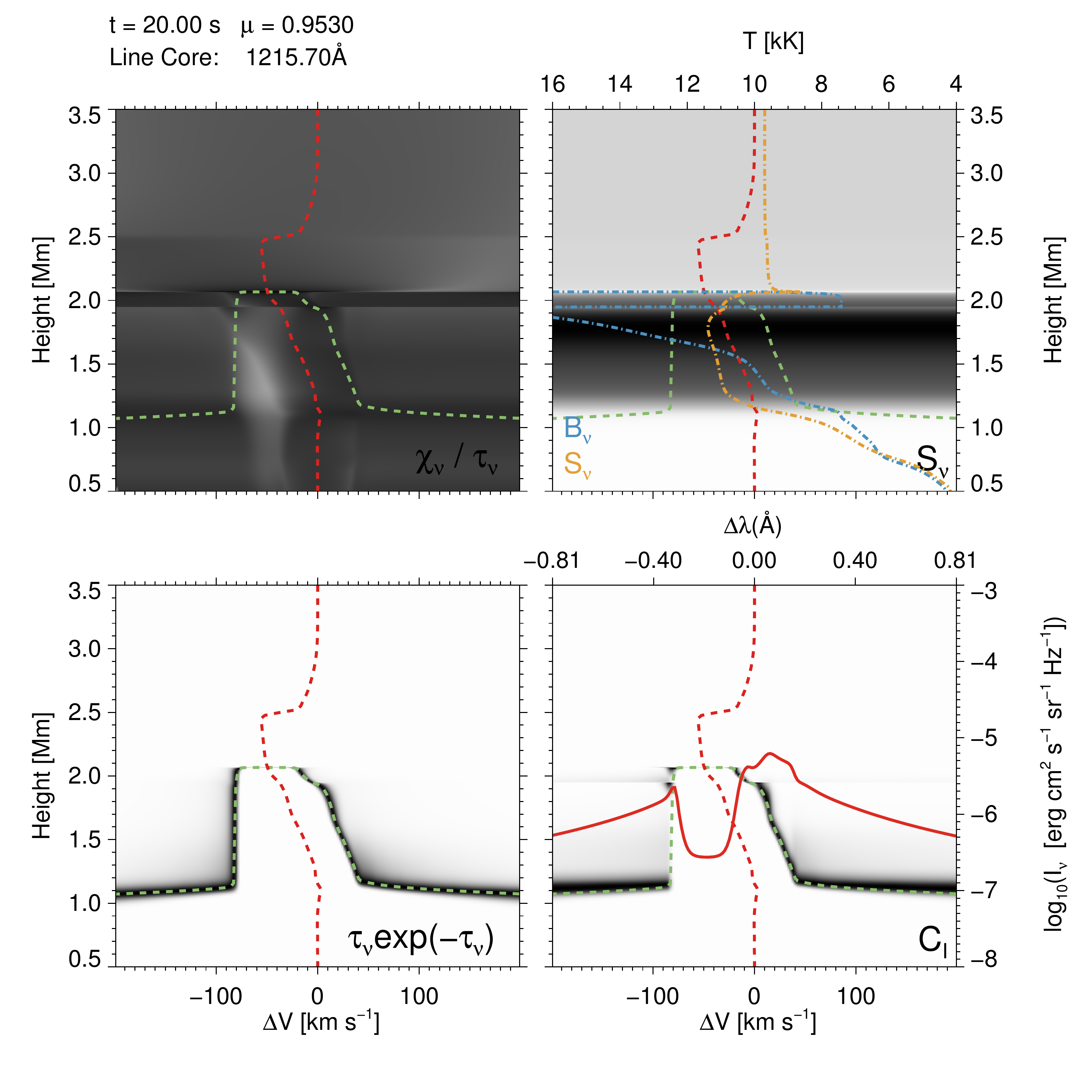}{0.45\textwidth}{(a)}
		\fig{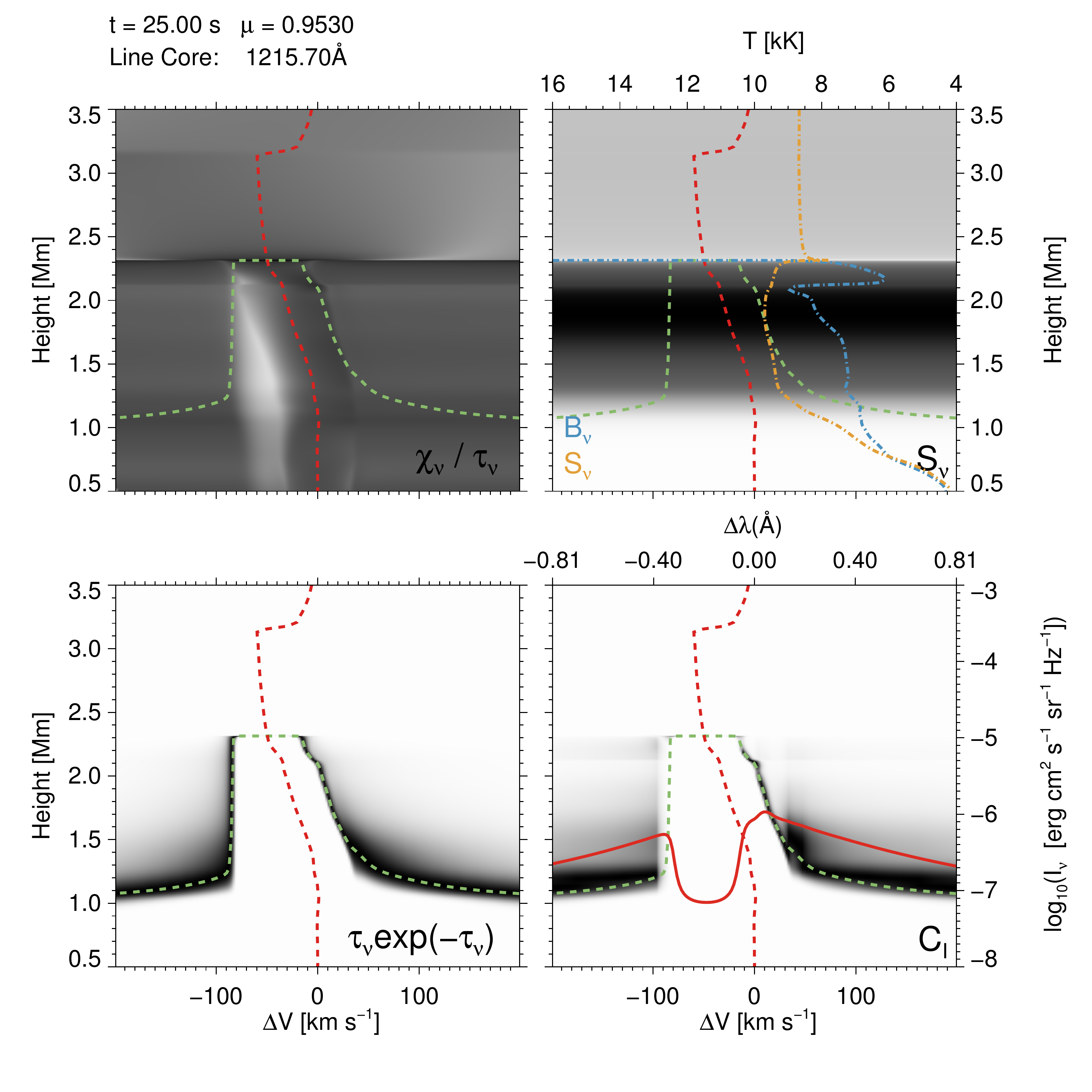}{0.45\textwidth}{(b)}}
	
	\caption{Line contribution functions for Ly-$\alpha$ at two times during the F10D8 simulation. All lines and panels retain their meanings from Figure \ref{Figure3}. \label{Figure9}}
\end{figure*}

We examine the formation of the Ly-$\alpha$ line in Figure \ref{Figure9} at similar times as those during the F10D3 simulation. At t=20 s, the electron beam has just stopped heating the atmosphere, and has resulted in an upflow through much of the chromosphere. The atmosphere has a peak velocity of almost 60 km s$^{-1}$, but the velocity gradient above the transition region is steeper than in the F10D3 simulation as the coronal layers are almost stationary. As a result of the upflow, the $\tau_{\nu}$ = 1 surface is noticeably asymmetric as the opacity structure of the moving plasma is shifted to the blue as it is carried upwards. The height of this surface indicates that the line core is formed at z=2.05 Mm, whereas at the same time in the previous simulation it was formed higher.

The line contribution function shows that almost all of the line is optically-thick, with only small contributions from altitudes above the $\tau_{\nu}$ = 1 surface. $S_{\nu}$ again exhibits a maximum at an altitude below the formation height of the line core, meaning that the resulting line wings contain a greater amount of emission than the line core, and that the line profile is centrally reversed. The central reversal in Ly-$\alpha$ is heavily blueshifted, and is positioned entirely within the blue wing as a result of the core being formed in the presence of a 50 km s$^{-1}$ upflow.
	
Curiously, the Ly-$\alpha$ line at t=20 s in the F10D3 simulation also forms in the presence of an upflow of this speed, but the blueshift in its central reversal is not as pronounced as in this model. From Figure \ref{Figure3}a, it can be seen that the line core forms over a larger vertical extent in the F10D3 simulation compared to the F10D8 simulation. In the latter the line core forms in a very narrow layer. This perhaps suggests that the blueshift in the core of Ly-$\alpha$ in Figure \ref{Figure9}a is more indicative of the dynamics, since it samples less of the atmosphere.

At t=25 s (Figure \ref{Figure9}b), there is no longer any energy being injected and the atmosphere is relaxing. Plasma continues to be carried upwards by an atmospheric flow, which still exhibits a prominent velocity gradient at coronal heights, and maintains a blue asymmetry in the $\tau_{\nu}$ = 1 surface. The height of this surface has also increased, indicating the Ly-$\alpha$ line core now forms at a height of 2.3 Mm.

As before, $S_{\nu}$ peaks at an intermediate altitude between the core and wing formation heights, and has undergone an overall decrease since t=20 s. As a result, the emergent Ly-$\alpha$ line is centrally-reversed and less intense than at t=20 s. The line contribution function at t=25 s indicates a slightly larger fraction of optically-thin emission relative to t=20 s, which is mainly concentrated in the red wing. The upflow velocity at the core-formation height has not diminished, and so the central reversal still persists exclusively in the blue wing.

\subsubsection{Velocities from RADYN}

\begin{figure*} 
	\figurenum{10}
	\gridline{\fig{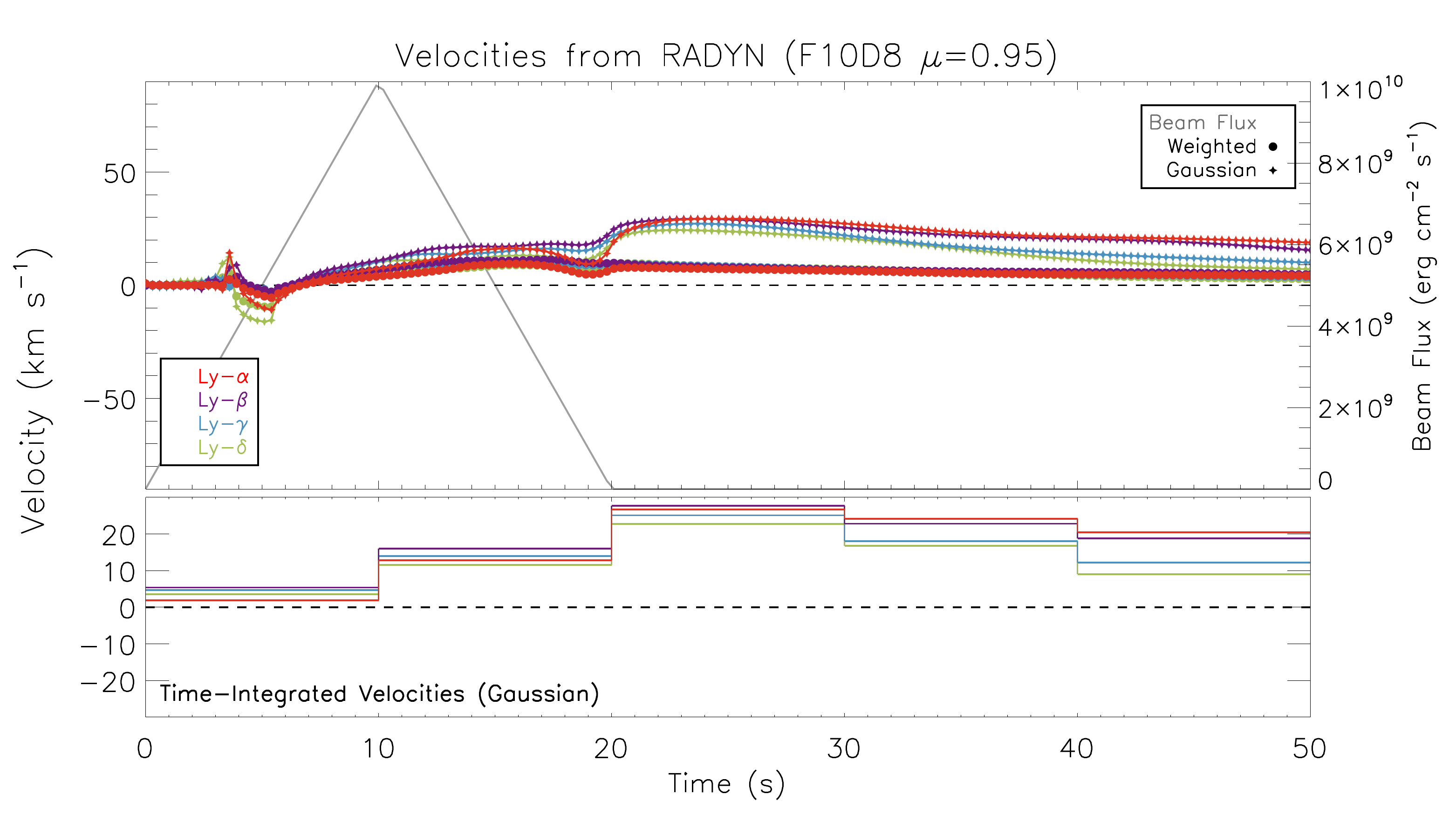}{0.56\textwidth}{(a)}
		\fig{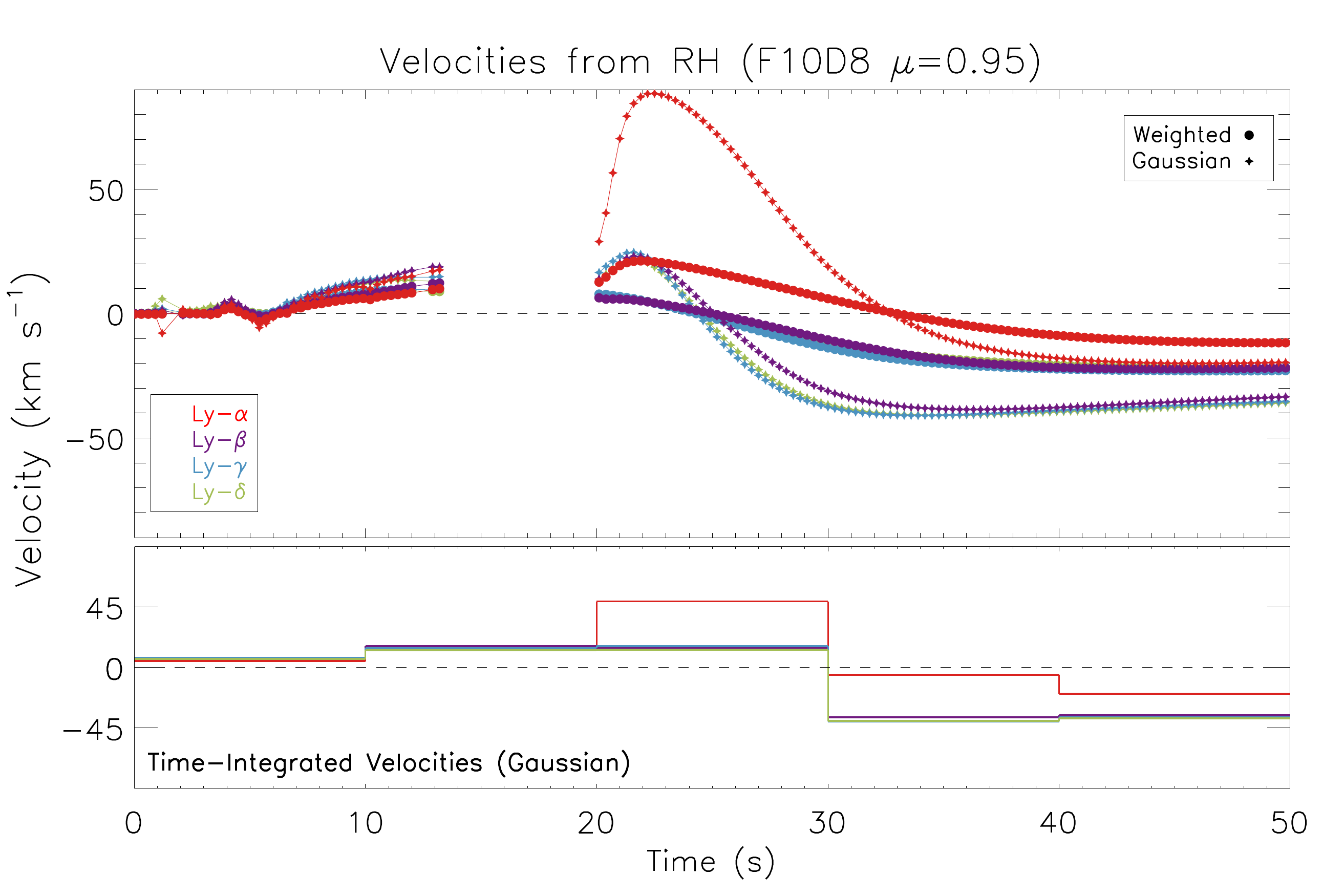}{0.47\textwidth}{(b)}}	
	\caption{Doppler velocities of the F10D8 line profiles from RADYN (a) and from passing the RADYN atmospheres through RH (b). Some of the atmospheres towards the end of the beam-heating stage did not converge in RH.  \label{Figure10}}
\end{figure*}

The Lyman lines from the F10D8 simulation are again convolved with EVE's instrumental profile, and Doppler shifts are calculated in the Ly-$\alpha$ through $\delta$ lines. Resulting velocity profiles, obtained from both the RADYN profiles and those after calculation in RH, are shown in Figure \ref{Figure10}. The velocities from RADYN (Figure \ref{Figure10}a) are dominated by persistent redshifted signatures in all of the Lyman lines throughout the entire duration of the simulation.

These redshifts suggest downflows of between 20-30 km s$^{-1}$. As in the F10D3 simulation, the source of these redshifts can be easily understood by visual inspection of the line profiles. From Figure \ref{Figure9}, it is clear that while the line cores are heavily blueshifted, they are also centrally reversed, and so when the profiles undergo convolution it is the red wing that becomes accentuated.

\subsubsection{Velocities from RH}

The velocity profiles obtained from the RH solutions (Figure \ref{Figure10}b) are more complex. Between t=10-20 s, RH fails to converge, likely as a result of the steep gradients in the atmosphere at these times. The velocity profiles throughout the first 10 s generally agree with those found from RADYN, with redshift signatures suggesting downflows of around 20 km s$^{-1}$, although the initial RH solutions for Ly-$\gamma$ and Ly-$\delta$ (t=1-2 s) are temporarily skewed by only half of each line profile being modelled correctly.

As observed in the F10D3 simulation, evolution of the line profiles after the beam-heating stops becomes significantly different with respect to the profiles computed from RADYN. As before, the higher order lines eventually transition into displaying blueshifts as the wing intensities in the line profiles diminish, allowing the blueshifts in the line core to be retained after the instrumental convolution. Ly-$\alpha$ however exhibits curious behaviour, with velocities from Gaussian fitting briefly suggesting downflow speeds of almost 100 km s$^{-1}$.

Shortly after the beam stops heating the atmosphere (t=20-25 s), the Ly-$\alpha$ line as computed from RH continues to display a central reversal (which is still blueshifted), but crucially does not have comparable intensities in the blue and red wings. In RADYN, the wings either side of the central reversal are typically similar in intensity. Conversely, in RH, the emergent blue wing of Ly-$\alpha$ shortly after the beam injection is notably less intense than the red wing. The combination of the blueshifted central reversal and lack of an appreciably intense blue wing acts to produce a line profile that has a very strong red asymmetry. This leads to the strongly redshifted signatures in Figure \ref{Figure10}b.

The F10D8 simulation shares similarities with its low-$\delta$ counterpart, namely the presence of clear blueshifts in the centrally-reversed line cores. As with the F10D3 simulation, these features are lost when the line profiles undergo convolution, and produce profiles with strong red asymmetries. Velocities at the core formation height are similar to those in the F10D3 simulation, but the altitude of this region is generally lower in the $\delta=3$ case. The velocity structure in the atmosphere is comparably more complex, consisting of a sharp gradient at higher altitudes, which is not observed in the F10D3 simulation. As in the F10D3 simulation, neither the direction or the extent of the atmospheric velocity is recovered by the line profiles after convolution.

\subsection{The 3F10D8 simulation}\label{subsec:3f10results}

\begin{figure*} 
	\figurenum{11}
	\plotone{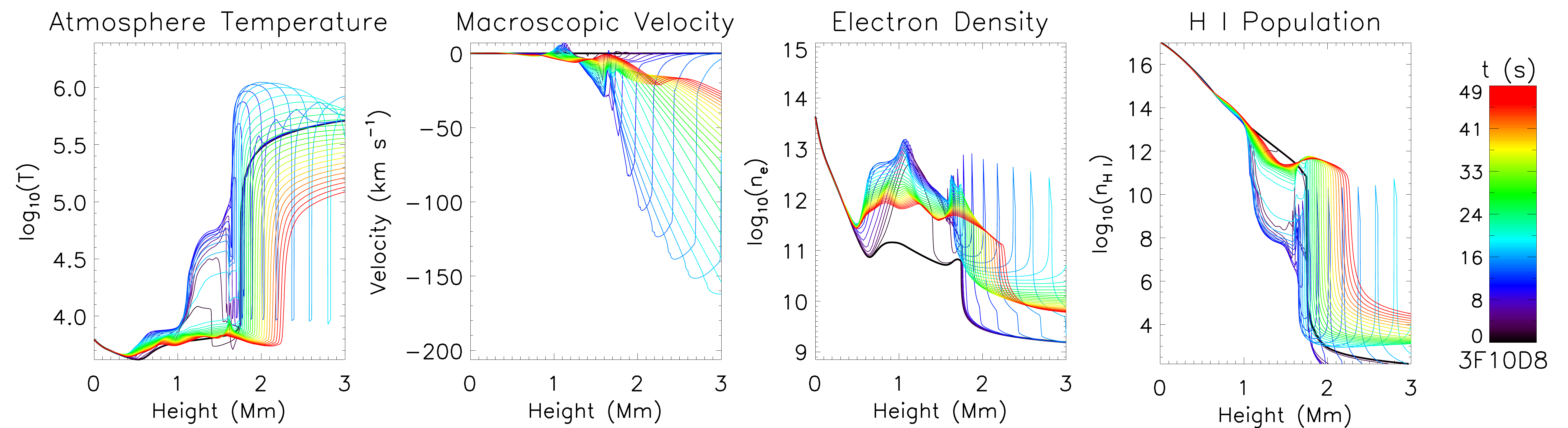}
	\caption{Atmospheric evolution during the 3F10D8 simulation. The pre-flare atmosphere is indicated by the thick black line. Quantities are plotted at $1.5$ s intervals. \label{Figure11}}
\end{figure*}

The overall beam flux in the 3F10D8 simulation ($\delta=8$, E$_{c}=25$ keV) is slightly enhanced above the previous two simulations. As with the F10D8 simulation, the high $\delta$ leads to the electrons being stopped higher in the chromosphere than the F10D3 simulation. The evolution of the atmospheric variables is shown in Figure \ref{Figure11}, in which we can see much more prominent spikes in electron density in the upper chromosphere compared to the F10D8 simulation.

 We also see that the plasma reaches much higher speeds than in both F10 simulations. We note a more complex, structured temperature profile, with a maximum of around T = 10$^{6}$ K being attained early in the simulation. Figure \ref{Figure11} shows that the altitude of the transition region briefly decreases during the beam injection, but moves upwards at later times.
 
The spikes in the electron density between t=10-20 s are concurrent and cospatial with relatively narrow but deep troughs in the atmospheric temperature profile and enhancements in the total population of neutral hydrogen (Figure \ref{Figure11}). These features move upwards with time, and could indicate a front of cool, dense material being pushed upwards as chromospheric evaporation takes place. The simultaneous enhancement in the total population of neutral hydrogen indicates that this is a propagation of mass, and not just an ionisation front.

\begin{figure*} 
	\figurenum{12}
	\gridline{\fig{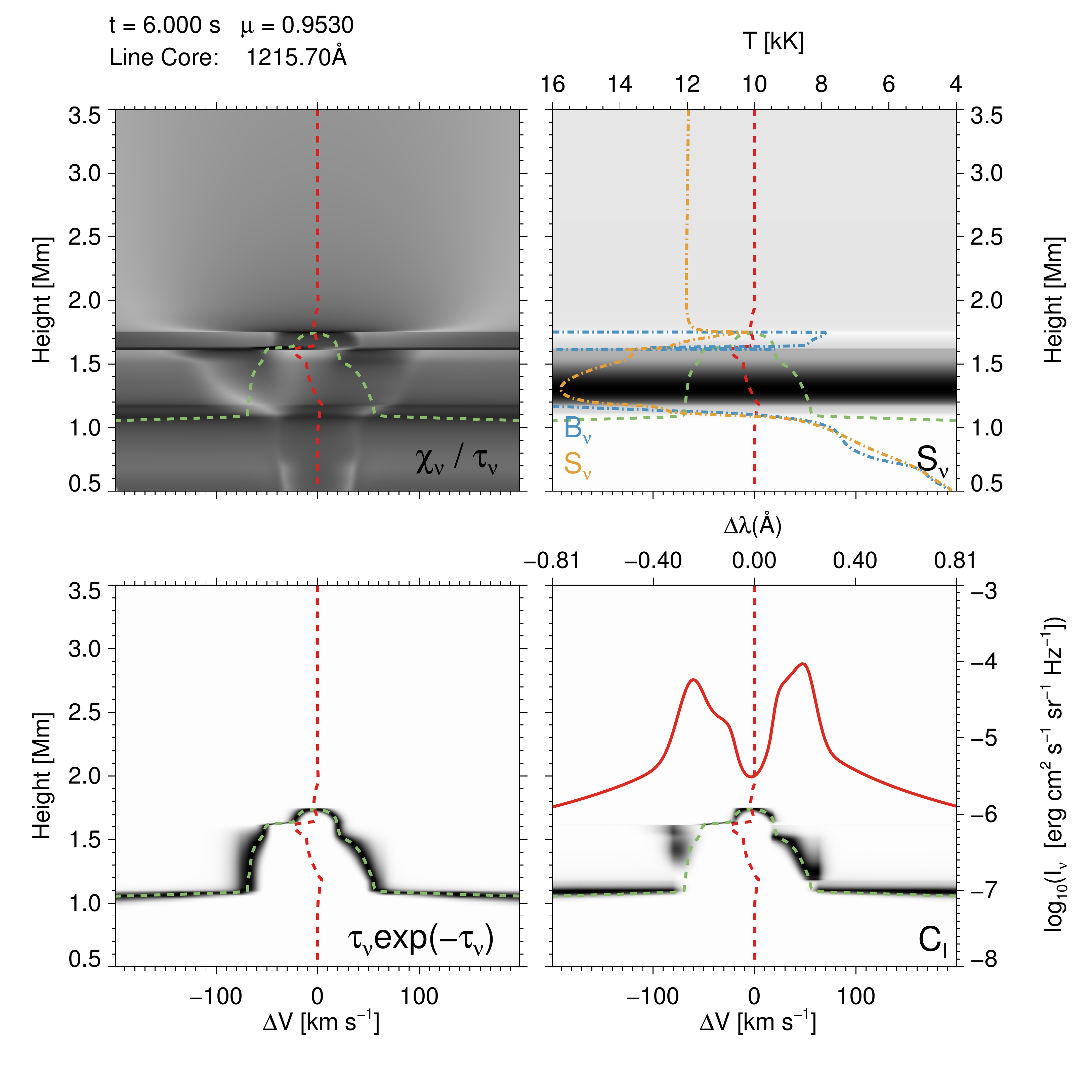}{0.45\textwidth}{(a)}
		\fig{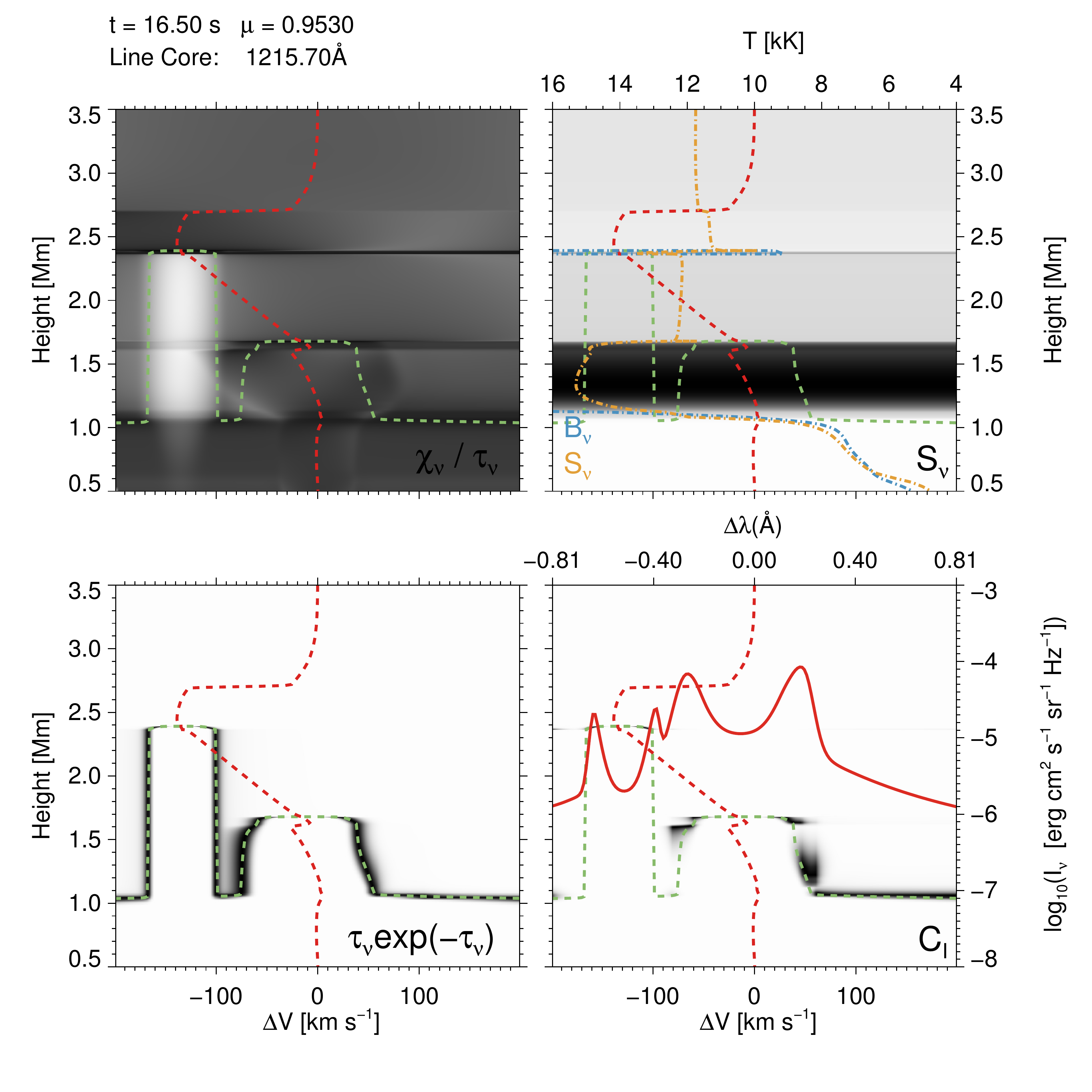}{0.45\textwidth}{(b)}}
	\gridline{\fig{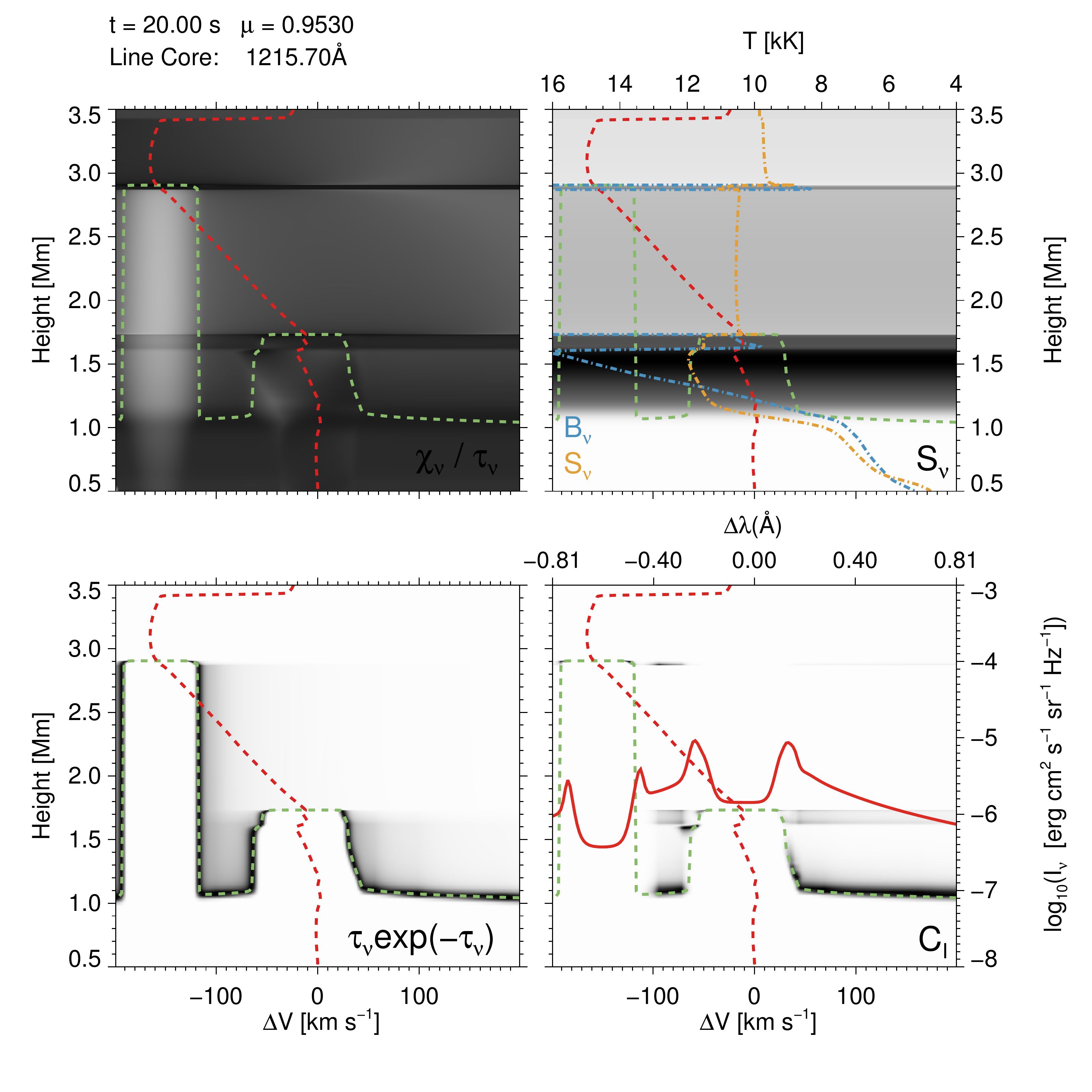}{0.45\textwidth}{(c)}
		\fig{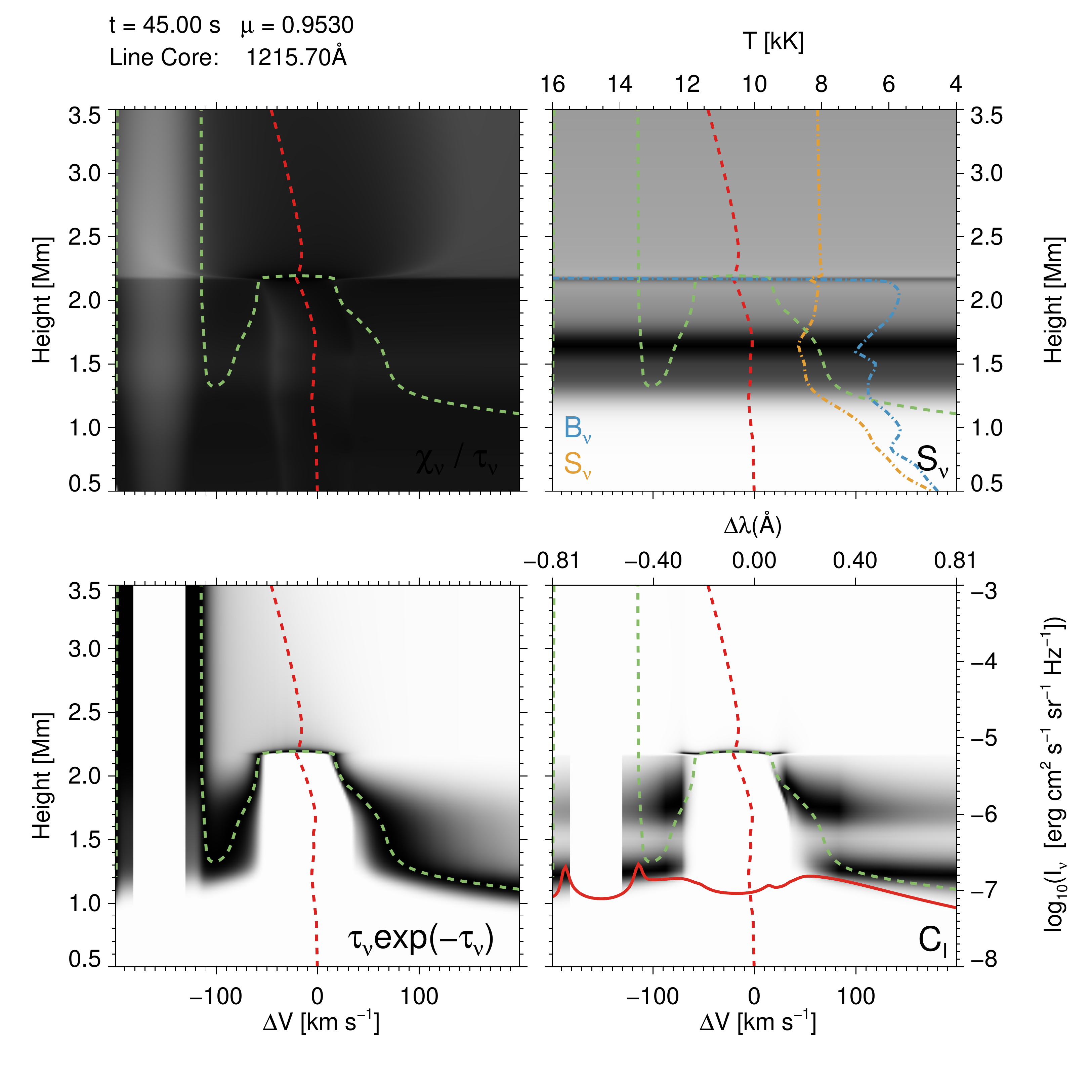}{0.45\textwidth}{(d)}}
	
	\caption{Line contribution functions for Ly-$\alpha$ throughout the duration of the 3F10D8 simulation. All lines and panels retain their meanings from Figure \ref{Figure3}. \label{Figure12}}
\end{figure*}

We investigate the properties of the line contribution function for Ly-$\alpha$ in Figure \ref{Figure12}. Because the higher-order Lyman lines behave similarly to Ly-$\alpha$, and for conciseness, we display only Ly-$\alpha$ in favour of presenting more timesteps in the simulation.

In Figure \ref{Figure12}a, we again observe similar line characteristics to those seen in Figure \ref{Figure3}, with an asymmetric $\tau_{\nu}$=1 surface and S$_{\nu}$ peaking at an intermediate height between the core and wing formation heights. The plasma is slightly upflowing at this point, and there appears to be complex structure within the central reversal. Much of the wing emission is optically thin, as indicated by the contributions to the intensity that appear above the $\tau_{\nu}$=1 surface.

Between the beam peak-time and the beam switching off, the line formation becomes much more complex. At t=16.5 s (Figure \ref{Figure12}b), the line source function still has a maximum below the core formation height, so the profile is still centrally reversed. A secondary feature of the $\tau_{\nu}$=1 surface has emerged. This feature originated from the original asymmetric surface close to the line core, and propagates through the blue wing, peaking at higher altitudes as a function of time. Furthermore, the maximum height of the secondary $\tau_{\nu}$=1 surface feature lies at the same height as the maximum velocity in the atmosphere, which has grown considerably since t=6 s to almost 150 km s$^{-1}$. 

This secondary feature in the $\tau_{\nu}$=1 surface acts to introduce an additional, highly blueshifted source of line emission which is linked in formation height to the peak speed of the atmosphere. This secondary component is predominantly optically-thick, mainly contributing emission at the same height as the $\tau_{\nu}$=1 surface. This additional source of emission is likely related to the cool, dense front revealed by the features seen in the temperature and density profiles in Figure \ref{Figure11}.

As the feature moves higher through the atmosphere (Figure \ref{Figure12}c), it propagates further through the blue wing of the line as the atmospheric velocity increases. The highly-blueshifted ``core" remains optically thick, while some additional optically thin wing emission is also produced by the dense upflow. By $t=20$ s, the source function has undergone an overall decrease, leading to a general reduction in emission from the line profile.

By t=45 s (Figure \ref{Figure12}d), the atmospheric velocity has weakened in the region of the chromosphere and is extended over a large range in height. The intensity of the line profile has diminished considerably due to the low magnitude of the source function over the line formation height, but the secondary blueshifted component persists and acts to strengthen the blue wing.

\subsubsection{Velocities from RADYN}\label{3f10d8velocitiesradyn}

\begin{figure*} 
	\figurenum{13}
	\gridline{\fig{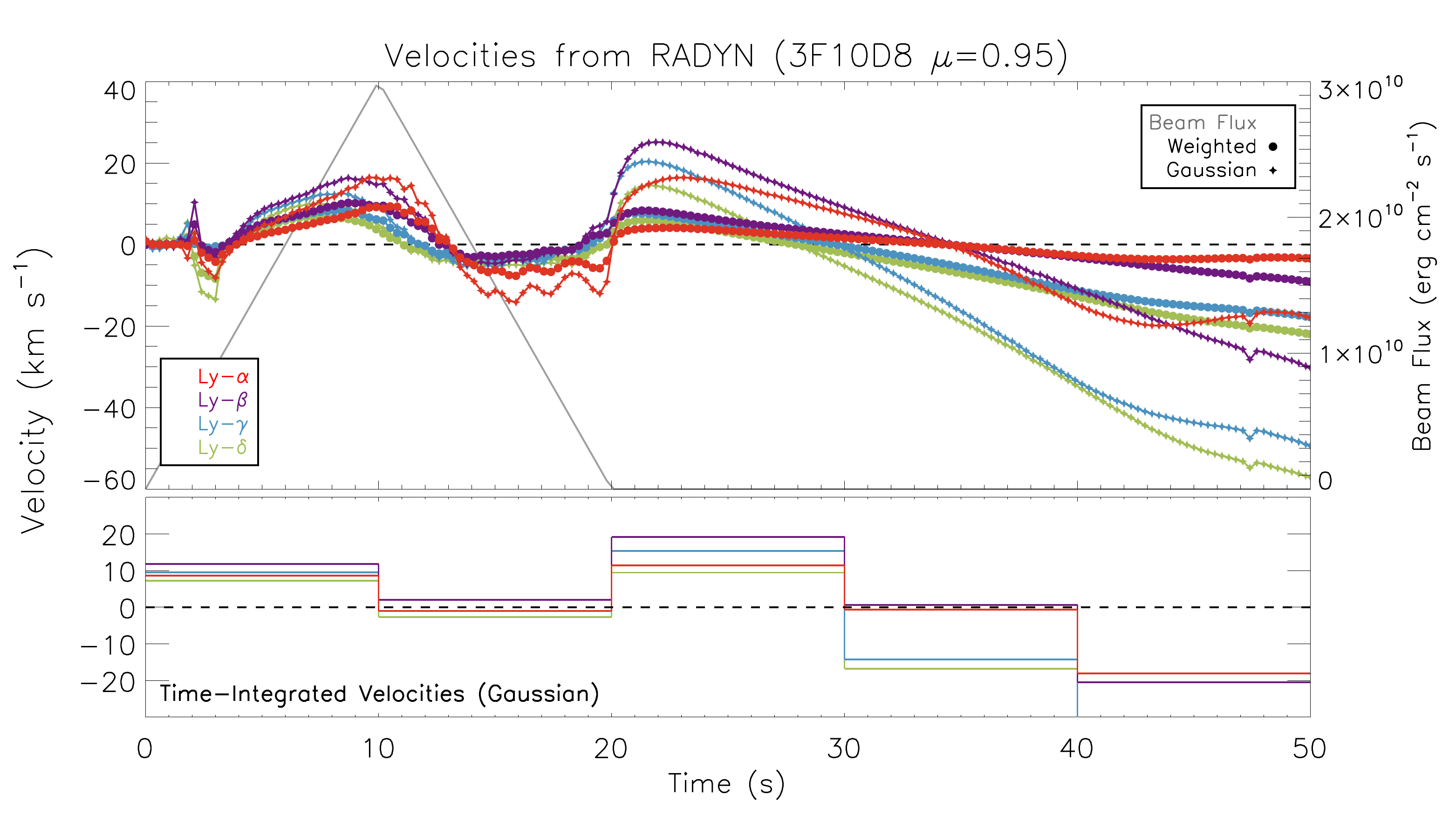}{0.56\textwidth}{(a)}
		\fig{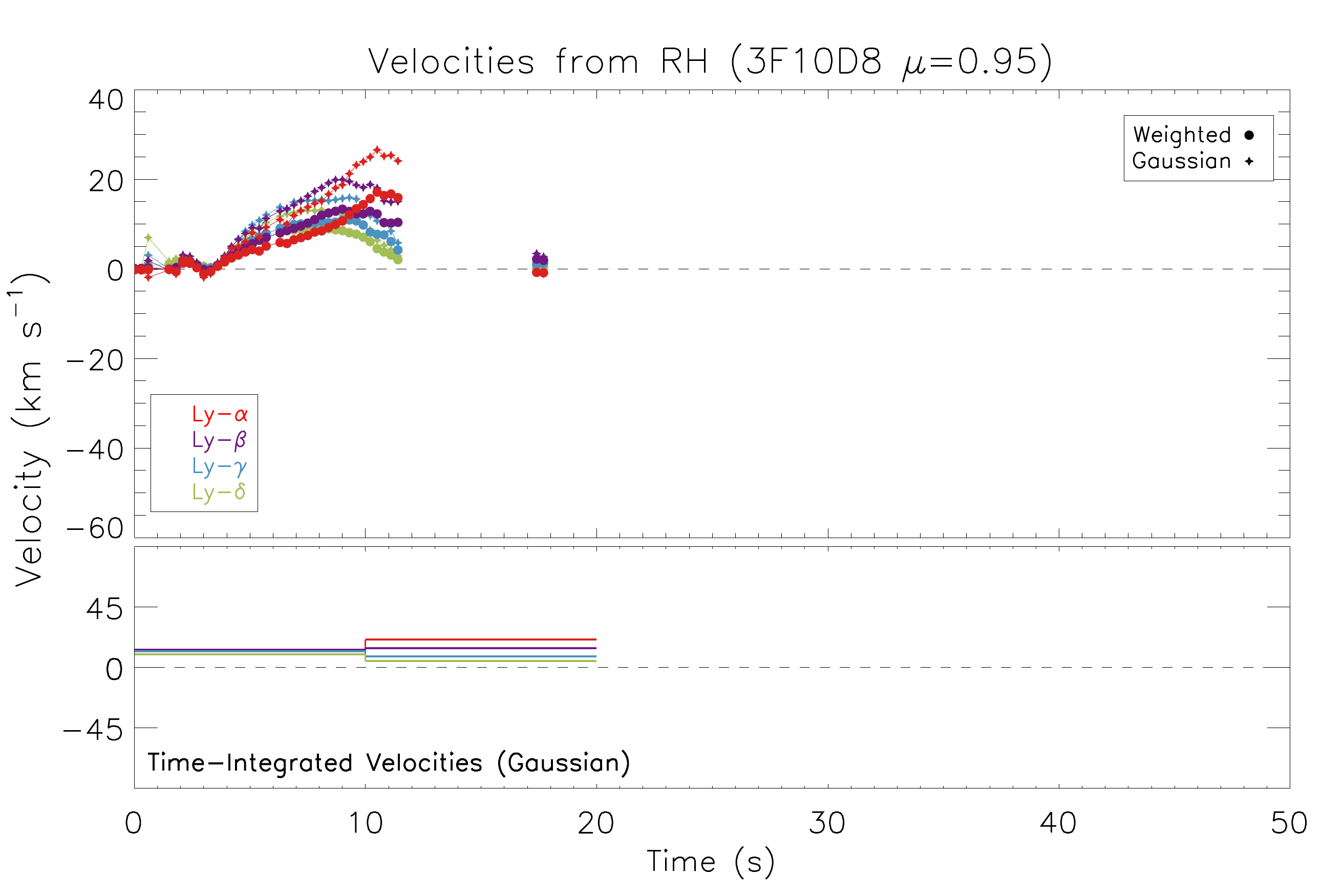}{0.47\textwidth}{(b)}}	
	\caption{Doppler velocities of the 3F10D8 line profiles from RADYN (a) and from passing the RADYN atmospheres through RH (b). Note that many atmospheres failed to converge in RH, and so only the initial part of the simulation is solved in RH.  \label{Figure13}}
\end{figure*}

Synthetic Doppler velocities are again calculated after simulating the EVE instrumental response and measuring the line centroid shifts. The velocities obtained from analysis of the RADYN and RH profiles obtained from the 3F10D8 simulation are plotted in Figure \ref{Figure13}. Due to the very steep gradients in the atmospheric structure in this simulation, many of the snapshots could not converge in the RH code and thus the results in \ref{Figure13}b do not constitute the entire length of the simulation as of that in \ref{Figure13}a.

The initial 10 seconds of the simulation display redshifts corresponding to downflows of around 15 km s$^{-1}$ in Ly-$\alpha$ and Ly-$\beta$, and 5-10 km s$^{-1}$ in the higher order lines. As with the F10D3 simulation, these perceived downflows are not due to redshifted emission, but are a consequence of the EVE instrumental profile masking a blueshifted core with an asymmetric red wing. The profiles after instrumental convolution are initially in absorption between t=0-3 s, transitioning into emission earlier than in the F10D3 simulation.

Between t=10-20 s, the effects of the dense upflow become apparent in the velocity profiles. While the secondary line component is self reversed, it also has narrow wing-like enhancements apparent in Figure \ref{Figure12}b which act to increase the overall amount of emission in the blue wing relative to the red wing. This feature counteracts the reduction in blue-wing emission from the primary central reversal, and leads to the observation of faint blueshifts in the line between t=10-20 s.

At t=20 s, the secondary line component has developed a deeper reversal. The primary central reversal also remains slightly blueshifted, and the combination of these two features leads to a greater excavation of emission in the blue-wing. This leads to the simulated observations once again obtaining redshifts of the order 10-20 km s$^{-1}$.

After t=20 s, the beam heating has stopped and the atmosphere is in the process of relaxing. The line profiles are now faint, but the secondary blue-wing component does persist to an extent, eventually contributing an overall enhancement in emission to the blue wing. Therefore, during this time the velocity profiles register blueshifts.

\subsubsection{Velocities from RH}

In Figure \ref{Figure13}b, the velocities obtained from passing the RADYN snapshots through the RH code are plotted. The structure of the atmosphere in this simulation is computationally difficult for RH to solve, and so only a subset of snapshots during the beam-heating stage converged. The first 10 seconds are well sampled, and display redshifts in the lines of the order 10-20 km s$^{-1}$ which agrees rather well with the RADYN velocity profile (Figure \ref{Figure13}a). 

 The presence of the high density upflow evidently causes problems for convergence, as only two snapshots at around t=17.5 s converge. These snapshots, however, suggest very weak redshifts in the lines. In Figure \ref{Figure13}a, we see very weak blueshifts at this time, so it is likely that we are again seeing the effect of the secondary line component acting to temporarily increase the amount of emission in the blue wing.

It is interesting to note that despite the brief (around t=12-18 s) excursion to blueshifts due to the secondary line component, the velocity profiles again register redshifts throughout the first 10 seconds as a result of the centrally-reversed line cores being blueshifted. Maximum flow velocities of around 60 km s$^{-1}$ are observed at late times in the higher-order Ly-$\delta$ and Ly-$\gamma$ lines, with speeds across the lower-order Lyman series remaining around the 20 km s$^{-1}$ level. The diminishing of flows between t=10-20 s, and high upflow velocities at late times are both caused by the appearance and persistence of the blue-wing feature observed in Figure \ref{Figure12}. 

It is worth noting that the production of the secondary line component in this simulation is reliant both on the increased $F$ and $\delta$ values, as similar features are not found in lower-flux simulations with the same $\delta$ (F10D8 model), or in higher-flux simulations with a lower $\delta$ (F11D3 model).

\subsection{The F11D3 simulation}\label{subsec:f11results}

\begin{figure*} 
	\figurenum{14}
	\plotone{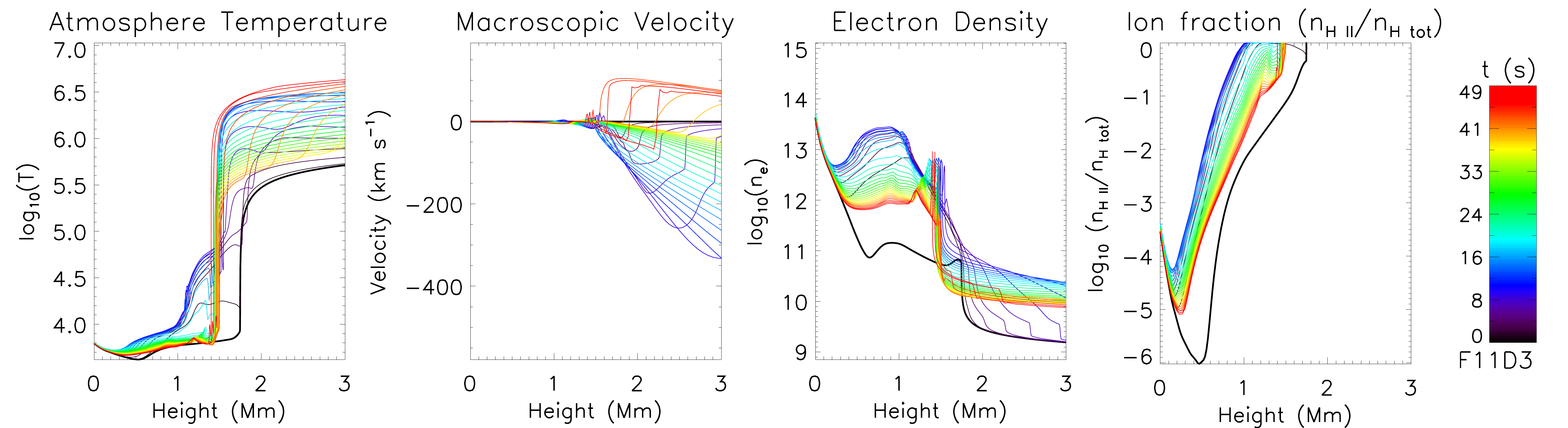}
	\caption{Atmospheric evolution during the F11D3 simulation. The pre-flare atmosphere is indicated by the thick black line. Quantities are plotted at $1.5$ s intervals. A complex velocity structure can be seen close to the transition region at t= $46$ s (in red). \label{Figure14}}
\end{figure*}

The final simulation in our study is a high-flux, hard beam ($\delta=3$, E$_{c}=25$ keV). Like the F10D3 simulation, this beam deposits a larger fraction of its energy in the lower chromosphere compared to the $\delta$=8 simulations. The atmospheric evolution is shown in Figure \ref{Figure14}, in which we see a prominent spike in the lower chromospheric temperature shortly after the beam is injected along with a brief compression of the chromosphere as the transition region makes a small excursion to lower altitudes. The transition region does not return to its initial altitude by the end of our simulation. The atmospheric temperature during beam heating is higher than that in the F10D3 simulation, attaining temperatures of almost T=10$^{6.5}$  K compared to T=10$^{5.9}$  K in the F10D3 simulation.

The flows initiated in this simulation are of a much greater magnitude than in any of the other simulations, with upflows attaining speeds of almost 400 km s$^{-1}$ at z=3 Mm. The sharp temperature boundary at t=50 s indicates that the transition region settles at an altitude of 1.4 Mm.

Near the end of the simulation (t=45-50 s), a sharp feature of enhanced electron density appears close to the transition region. At this time, we also observe a downflow in the atmospheric velocity of around 100 km s$^{-1}$. It is possible that the presence of downflowing material results in a compression of this region, which would enhance the electron density.

\begin{figure*} 
	\figurenum{15}
	\gridline{\fig{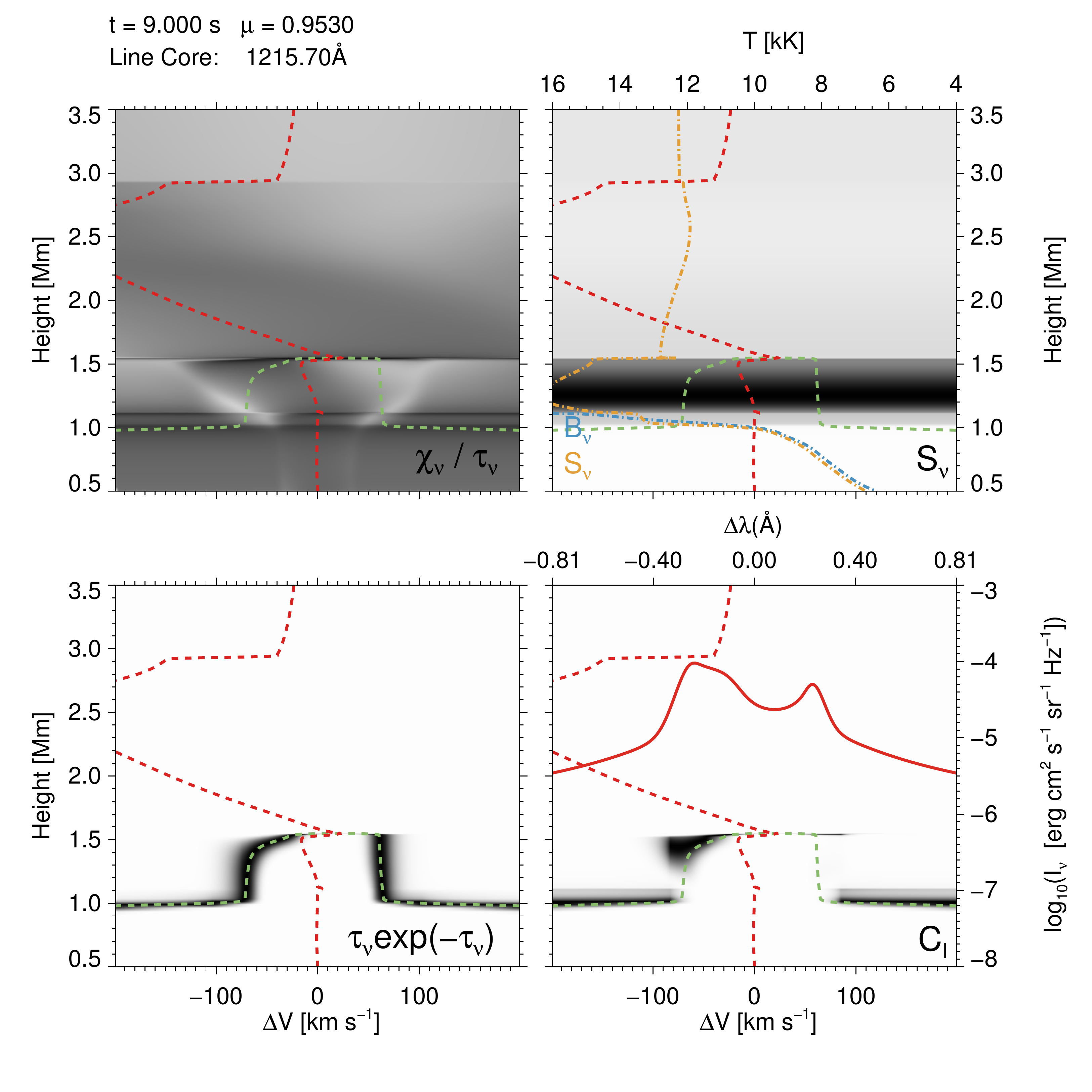}{0.45\textwidth}{(a)}
		\fig{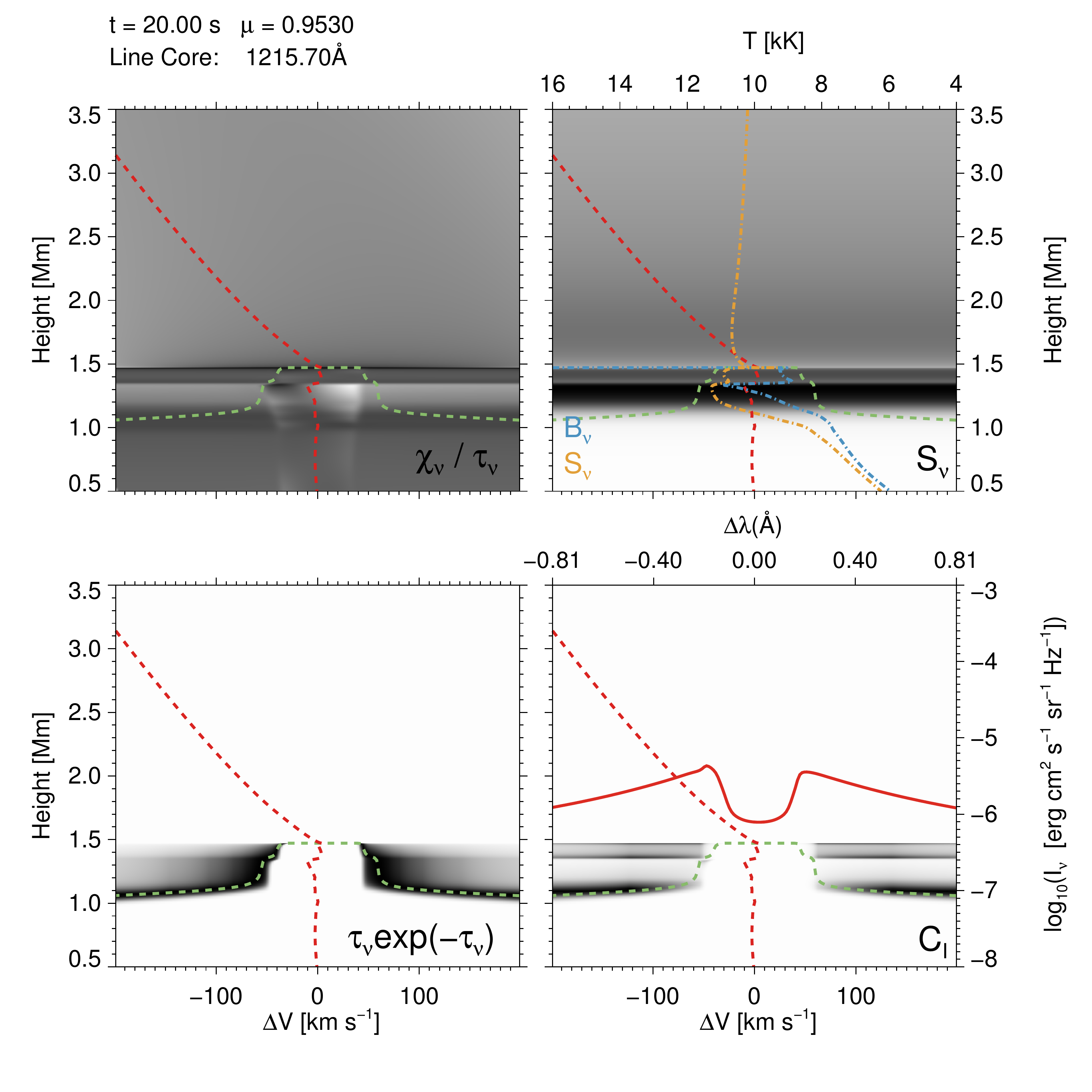}{0.45\textwidth}{(b)}}
	\gridline{\fig{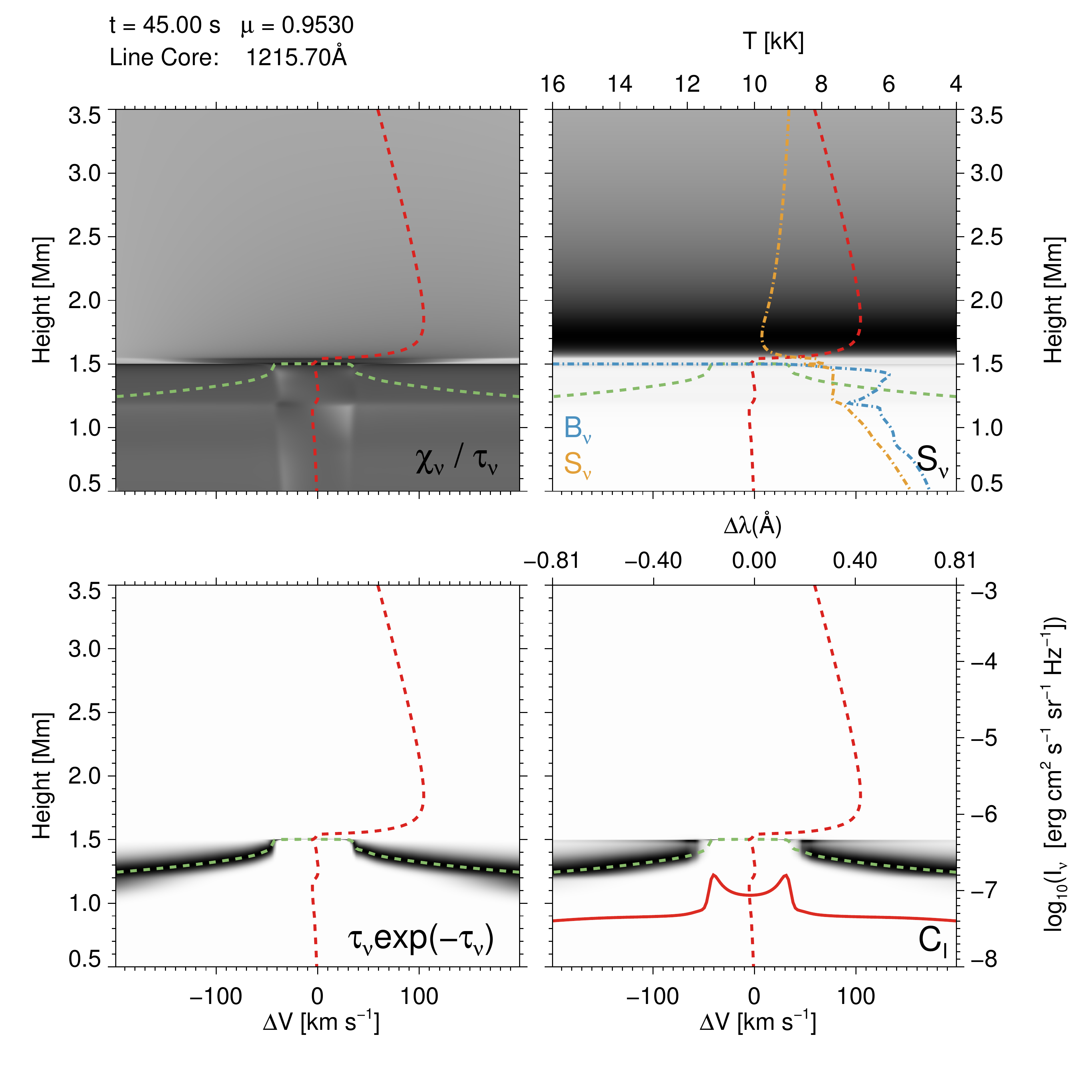}{0.45\textwidth}{(c)}
		\fig{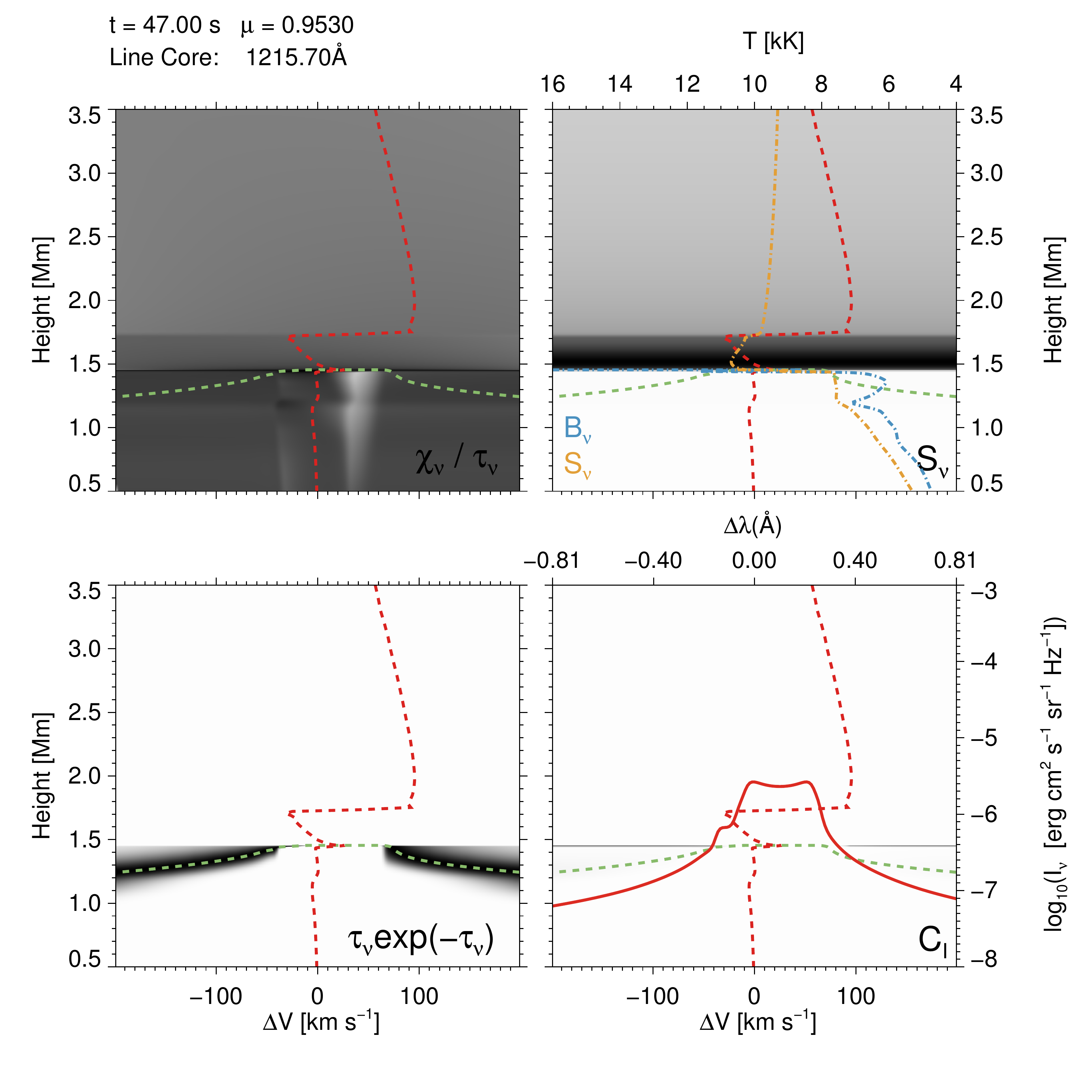}{0.45\textwidth}{(d)}}
	
	\caption{Line contribution functions for Ly-$\alpha$ throughout the duration of the F11D3 simulation.  \label{Figure15}}
\end{figure*}

The contribution functions for Ly-$\alpha$ at varying simulation times are plotted in Figure \ref{Figure15}. Again, because the other Lyman lines exhibit similar behaviour in this simulation, we forgo plotting higher order lines in favour of presenting more timesteps.

Figures \ref{Figure15}a and \ref{Figure15}b reveal that while the transition region and corona are subject to a very fast ($v>$  200 km s${-1}$) upflow, the height of core formation (z=1.5 Mm) is stationary. This was also found at the core-formation height of the \ion{Na}{1} D$_{1}$ in an F11 simulation by \citet{Kuridze2016}. Because of this, the line core is not Doppler shifted, and is centrally-reversed as $S_{\nu}$ peaks deeper in the atmosphere. At t=9 s, upflows reaching 20 km s$^{-1}$ are present below the core-formation height, with a very steep velocity gradient above. An enhancement in $\frac{\chi_{\nu}}{\tau_{\nu}}$ can be seen at this time in the blue wing, as a result of these upflows leading to an asymmetry in the $\tau_{\nu}=1$ surface. Curiously, a slight redshift is observed in the centrally-reversed line core at t=20 s, despite the lack of any downflow signature in the atmospheric velocity. At this time, the blue wing is also slightly more intense than the red wing, as a result of an enhancement in $\frac{\chi_{\nu}}{\tau_{\nu}}$.

At t=45 s, the atmosphere is in the process of relaxation (Figure \ref{Figure15}c). The $\tau_{\nu}$=1 surface has regained its symmetry and all parts of the line are in emission due to the source function peaking slightly above the core-formation height, which is very close to the transition region. Additionally, the line is largely symmetric, as it is not formed in the presence of any appreciable flows in the atmosphere. At this time, a downflow with a velocity of around $100$ km s$^{-1}$ can be seen propagating down from the corona.

This downflow reaches the core-formation height, and abruptly changes direction as if rebounding. This can be seen in Figure \ref{Figure15}d, and it clearly has an effect on the Ly-$\alpha$ line. The line source function is largely concentrated in the wake of the now-upflowing plasma, and a slight blue-wing enhancement can be seen in $\frac{\chi_{\nu}}{\tau_{\nu}}$. The line contribution function indicates that the emission is produced in an extremely thin region cospatial with the altitude from which the atmospheric flow changed direction. The resulting line profile is complex, with a dominant red-wing enhancement as a result of the core-formation height undergoing downflow. This is accompanied by a lesser contribution in the blue wing, likely produced by the rebounding upflow. The line is clearly more intense than at t=45 s, indicating that the complex dynamics at this time act to produce heightened levels of emission.

\begin{figure*} 
	\figurenum{16}
	\includegraphics[width=18.2cm]{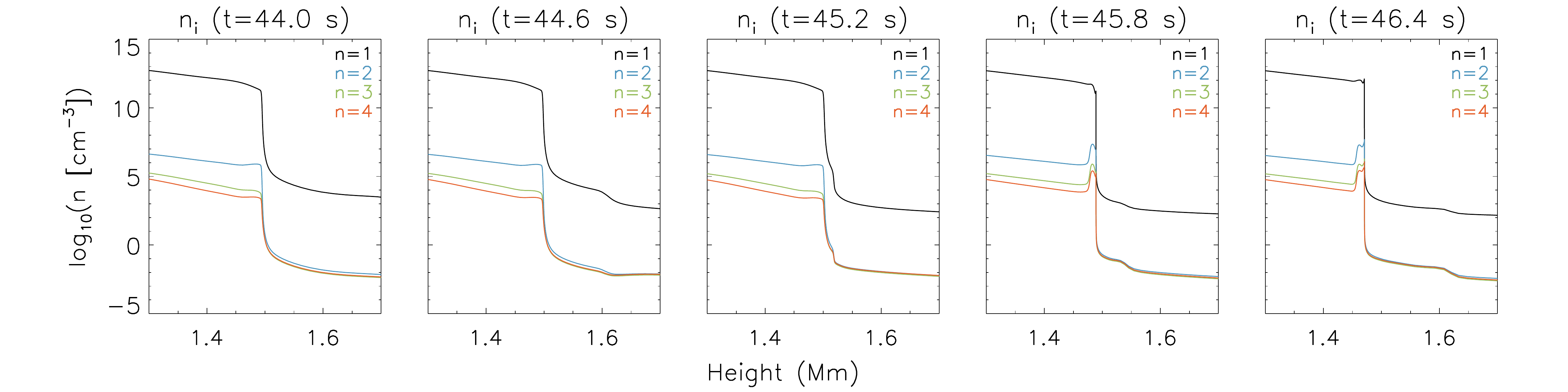}
	\caption{Level populations for hydrogen in the chromosphere around the time corresponding to the abrupt appearance of emission in the line wings (t=45 s) during the F11D3 simulation. \label{Figure16}}
\end{figure*}

 The level populations for hydrogen are plotted in Figure \ref{Figure16} for a series of timesteps around this feature, and a marked increase is seen in all levels at the same time as the appearance of the wing emission. 

Given the population enhancements and the rapid change in direction of the atmospheric velocity just above the $\tau_{\nu}$=1 surface, it is possible that a fast downward flow results in an increase in the local plasma density, allowing for an increase in the amount of collisional excitation at this height and population of the upper levels. This then leads to emission by de-excitation. A fraction of this emission may be blueshifted as the material is carried upwards by the ``rebounding" upflow.

\subsubsection{Velocities from RADYN}

\begin{figure*} 
	\figurenum{17}
	\gridline{\fig{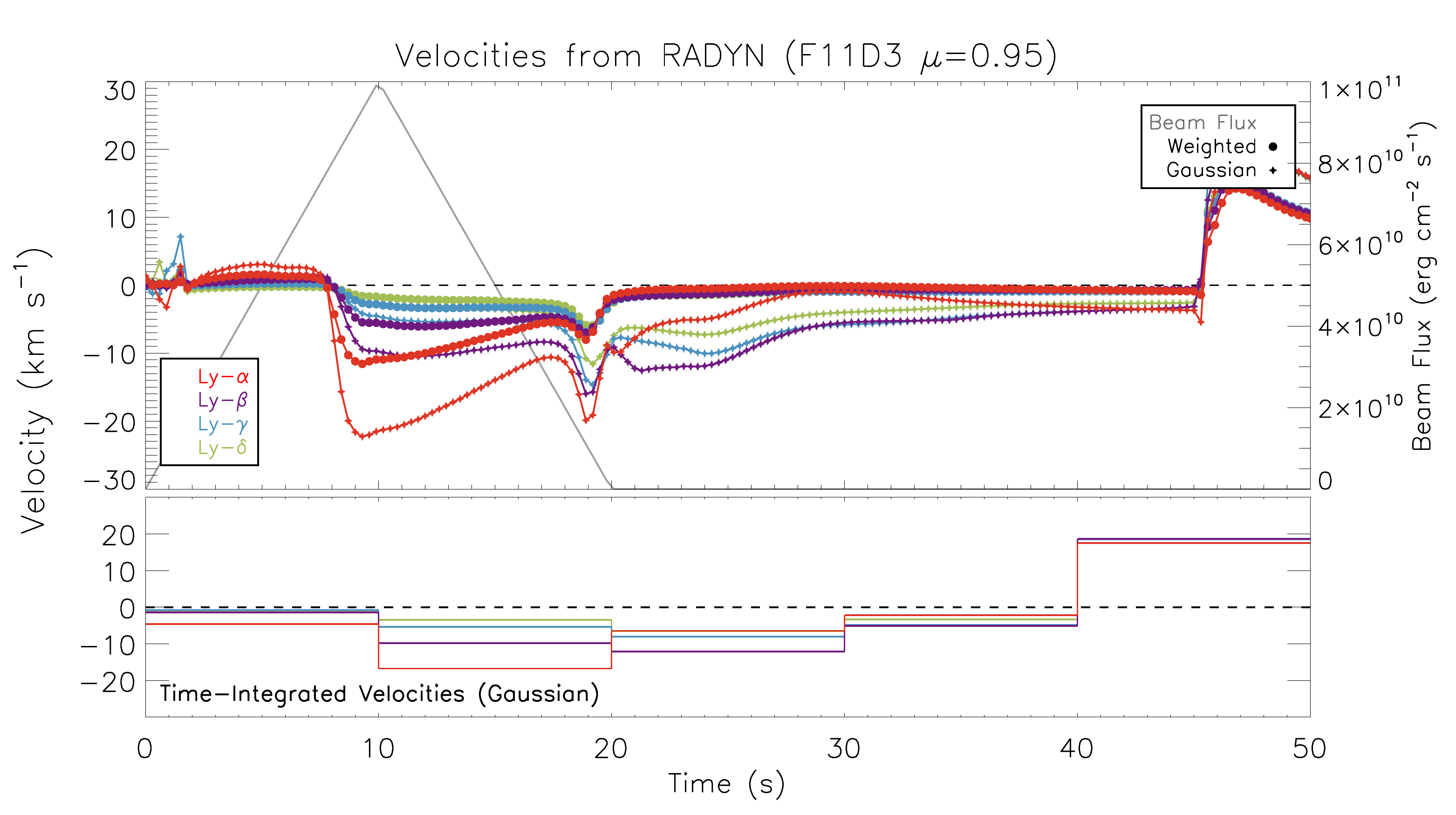}{0.56\textwidth}{(a)}
		\fig{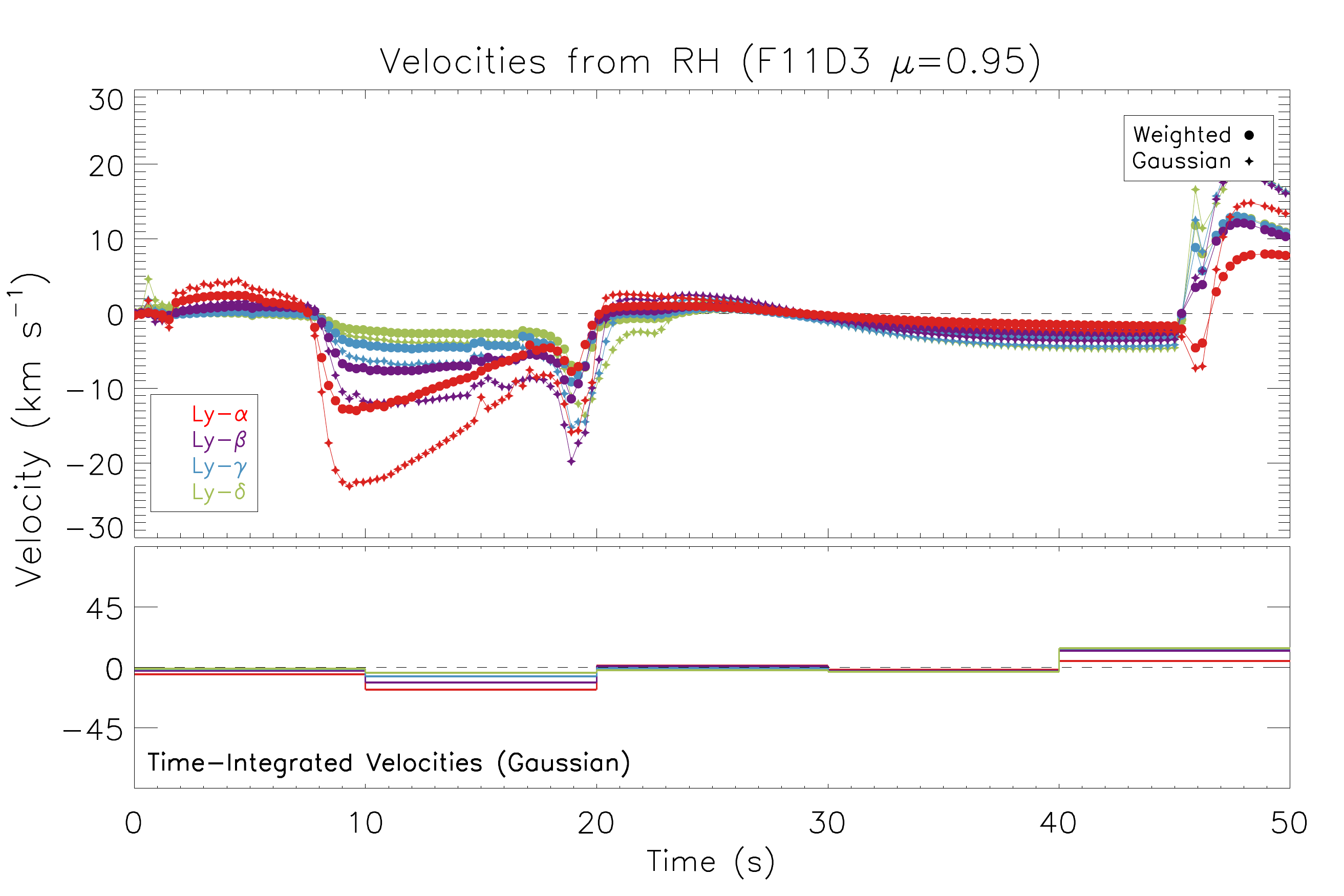}{0.47\textwidth}{(b)}}
	
	\caption{Doppler velocities of the F11D3 line profiles from RADYN (a) and from passing the RADYN atmospheres through RH (b).  \label{Figure17}}
\end{figure*}

The simulated Doppler velocity profiles for the F11D3 simulation are plotted in Figure \ref{Figure17}, again for both the RADYN and RH output. The first 20 seconds in the RADYN velocity profile (Figure \ref{Figure17}a) are primarily dominated by blueshifts. The line contribution functions for Ly-$\alpha$ at t=9 and t=20 s (Figures \ref{Figure15}a and \ref{Figure15}b) show profiles with shallow central reversals, but with heightened amounts of emission in the blue wing relative to the red wing. These are directly linked to the upflow in the atmosphere. While the blue wing is less pronounced at t=20 s, it is accompanied by a slight redshift in the centrally-reversed core, which acts to further accentuate the emission in the blue wing. These factors lead to blueshifts being registered in the velocity profiles throughout the beam-heating stage, which continue throughout the majority of the simulation. As before, Ly-$\alpha$ and Ly-$\beta$ provide the more pronounced signatures, suggesting upflow speeds of $23$ and $10$ km s$^{-1}$ respectively when obtained from Gaussian fitting.

Around t=45 s, all Lyman lines abruptly transition into displaying redshifts, with Ly-$\alpha$ indicating downflows of 20 km s$^{-1}$. As a result of the change in direction of the atmospheric flow described earlier, it can be seen from Figure \ref{Figure15}d that the Ly-$\alpha$ line is strongly redshifted as a result of the core-formation height undergoing downflow. As a result of this, the velocity profiles produce redshifts for the remainder of the simulation. For this simulation, the flows are mostly linked to emitting features, and as a result the synthetic Doppler velocity profiles generally match the genuine flow direction in the atmosphere. It should be noted that the redshift in the centrally-reversed line cores visible at t=20 s will partially contribute to the observed blueshifts in the velocity profiles at this time.

\subsubsection{Velocities from RH}

The velocities from RH (Figure \ref{Figure17}b) closely match those obtained from RADYN. Blueshifts are again observed throughout the beam-heating stage, with Ly-$\alpha$ and Ly-$\beta$ exhibiting the fastest speeds of 25 and 12 km s$^{-1 }$ respectively. Ly-$\gamma$ and Ly-$\delta$ suggest only weak upflows of around 5 km s$^{-1}$. Throughout the intermediate time (t=20-45 s), little to no Doppler shifts are measured in the lines, whereas the RADYN profiles show sustained blueshifted signatures. This is not dissimilar to the post-beam phase in the F10D3 simulation, where the wing intensities for the RADYN and RH profiles evolved differently and led to differences in the velocity profiles for the two codes (\textsection \ref{subsubsec:rhvelf10}).
 
Between t=45-50 s, the RH velocity profiles again echo those observed from the RADYN line profiles, with each of the Lyman lines transitioning into exhibiting redshifts. In RH, however, the Ly-$\alpha$ response to the change in direction of the atmospheric flow is less pronounced than in the higher order lines, which display redshifts corresponding to downflows of 20 km s$^{-1}$.

In summary, the F11D3 simulation remains dominated by blueshifts in the Lyman lines while the beam is injected, and primarily originate from emitting features in the line profiles, although this is bolstered by a slight redshift in the centrally-reversed line cores close to t=20 s. These blueshifted signatures transition into redshift when the downflowing plasma interacts with the core-formation height, which produces heightened levels of emission with distinctly redshifted profiles.

In all simulations, the Lyman lines are capable of indicating the motion of plasma upflowing through the atmosphere. Crucially, this behaviour is not recovered in simulated observations if the line profiles are centrally reversed (as in the F10 simulations), as the absorbing nature of the blueshifted line core acts to reduce the amount of emission in the blue wing rather than enhance it. \citet{Kuridze2015} also found that the red asymmetry in H$\alpha$ during the early stage of their F11 simulation was a result of the blueshift of the centrally-reversed line core. These results show that caution must be taken in associating line asymmetries to flows in the same direction.

\section{Simulation Of SPICE Profiles}\label{sec:spice}

The Solar Orbiter satellite will accommodate the SPICE (Spectral Investigation of the Coronal Environment) instrument \citep{Fludra2013}. SPICE covers two EUV wavelength bands, one of which will include the Ly-$\beta$ line. While we do not attempt to perform an exhaustive analysis of SPICE's future capability regarding Ly-$\beta$ observations, we do explore some of the basic concepts.

Much of this paper has focused on emulating the degradation of detailed model line profiles via the EVE instrument. Here, we perform a similar analysis but with SPICE's design parameters. The long wavelength (LW) band will cover the wavelength range $ 97.3$ nm $< \lambda < 104.9$ nm. The spectrograph will disperse sunlight onto the detector at a resolution of 0.0083 ~nm per pixel at 101 nm, and the line spread function will have an extent of 4 pixels, corresponding to a FWHM of around 0.04 nm about the Ly-$\beta$ line \citep{Fludra2013}.

\begin{figure*} 
	\figurenum{18}
	\plotone{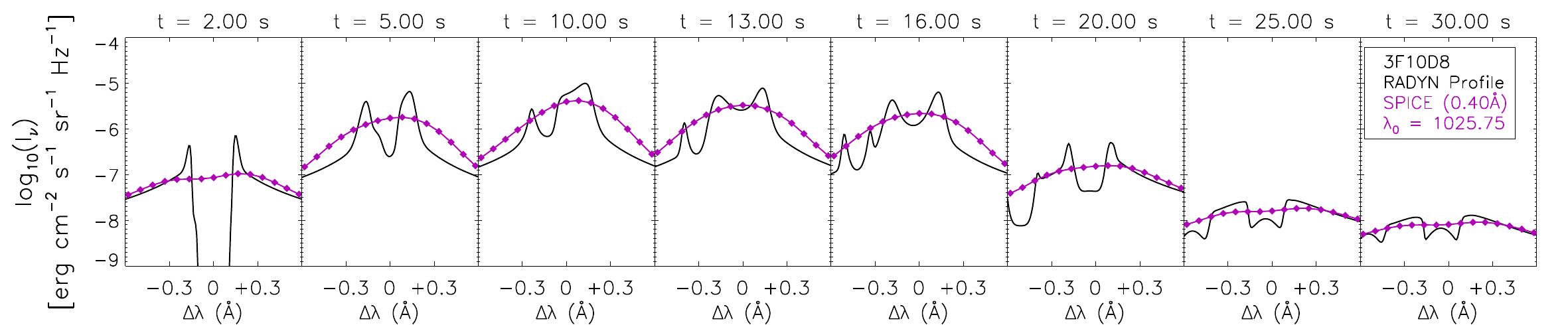}{(a)}
	\plotone{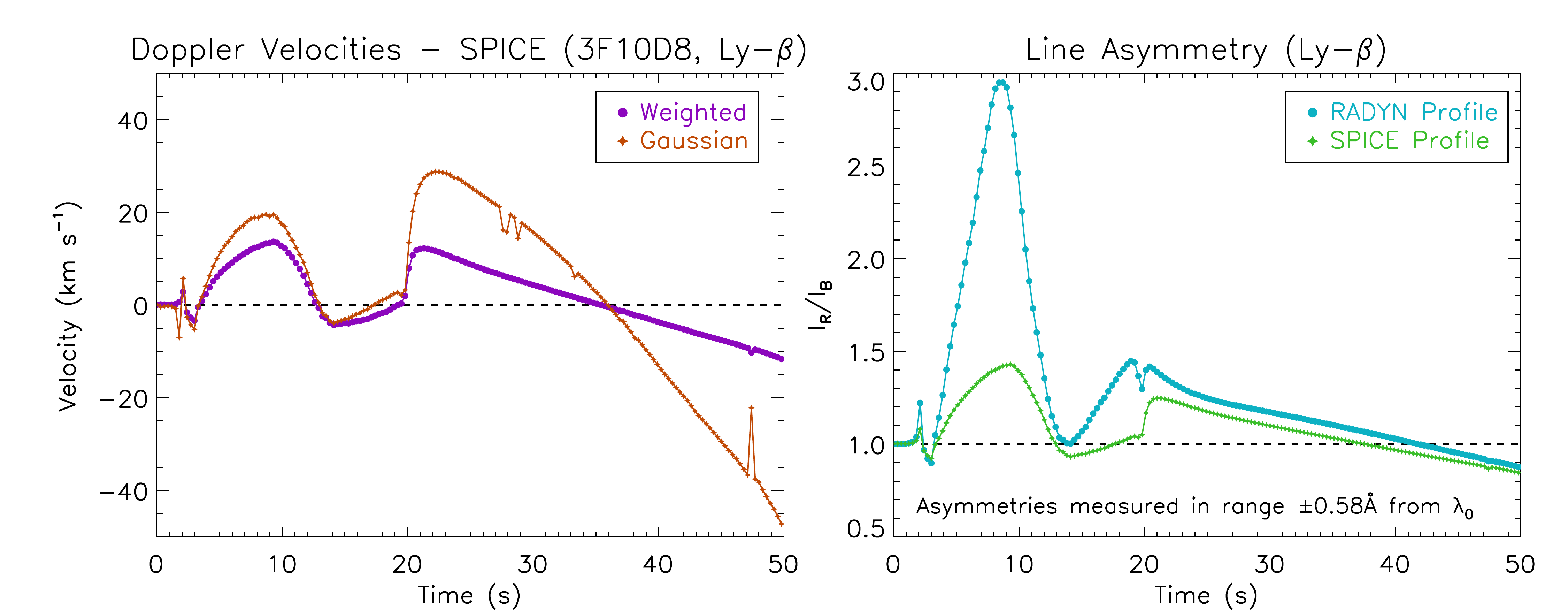}{(b)}
	
	\caption{Snapshots of the Ly-$\beta$ line during the 3F10D8 simulation with SPICE instrumentation effects (a) and the Doppler velocities obtained from measuring the line centroid variations of the degraded line profile, plotted alongside a running measurement of the line asymmetry. The line asymmetry is quantified by $\frac{I_{R}}{I_{B}}$, and is shown for both the RADYN output (blue) and for the convolved profiles (green). \label{Figure18}}
\end{figure*}

Given these parameters, we convolve the  Ly-$\beta$ profiles from RADYN with a Gaussian with a FWHM of 0.04 nm and then rebin the resulting convolved profiles to a wavelength spacing of 0.0083 nm per bin. We believe that this should reasonably approximate the effects of SPICE's instrumentation on the line. Because the exposure time of the instrument can vary, we do not include the effects of time integration. In Figure \ref{Figure18}, we show the results of SPICE's instrumentation on the Ly-$\beta$ line profile and the resultant Doppler velocities for the 3F10D8 simulation, choosing this simulation because the resulting Ly-$\beta$ line is particularly complex.

In Figure \ref{Figure18}a, it is clear that the SPICE instrumentation allows the strengthened red wing in the line profile to be detected during the first 10 seconds. It is not capable of resolving the secondary blue peak in the profile (around $-0.5$ \AA) at t=16 s (or any detailed feature at any time). However, it does result in a relatively symmetric profile being produced around t = 13-25 s, after which the red wing again becomes dominant as the secondary component contributes less emission. At t=2 s, we note that the primary central reversal is also subtly hinted at in the degraded profile by a slight dip in emission at the line core. We find that across all simulations, SPICE may be capable of observing the central reversals at certain times in the evolution of the Ly-$\beta$ line. 

We find that the Doppler velocities obtained in Figure \ref{Figure18}b are not very different from those in Figure \ref{Figure13}a, but while the shape of the velocity profiles remain similar, the results from SPICE allow greater maximum speeds to be measured. This is unsurprising, as the more detailed profiles from the instrument allow for Doppler shifts to be detected at a finer resolution than from EVE. As in \textsection \ref{3f10d8velocitiesradyn}, strong blueshifts are found at late times as a result of the persistence of the secondary line component while the stationary line component has diminished in intensity.

 We define the line asymmetry ($A$) as the ratio of the emission in the red wing to that in the blue wing (Equation \ref{eq:asymmetry}). In Figure \ref{Figure18}b, a running measurement of the asymmetry in Ly-$\beta$ is shown for both the emergent line profile from RADYN and from that post convolution, and it can be seen that SPICE manages to reasonably detect the asymmetries in this line.
 
  \begin{equation}\label{eq:asymmetry}
  A=\frac{\sum_{\lambda=\lambda_{0}}^{\lambda_{0}+0.58\AA} I_{\lambda}}{\sum_{\lambda=\lambda_{0}}^{\lambda_{0}-0.58\AA} I_{\lambda}}
  \end{equation}

SPICE is capable of resolving the central reversal of Ly-$\beta$ at certain times during this simulation, and this does lead to deviations in velocity results at some points because the single Gaussian becomes a very poor fit. Nonetheless, it is encouraging to find that these interesting central reversals could be retained in SPICE data, although these features may appear weak and exhibit only shallow dips in the core intensity.

\begin{figure*} 
	\figurenum{19}
	\plotone{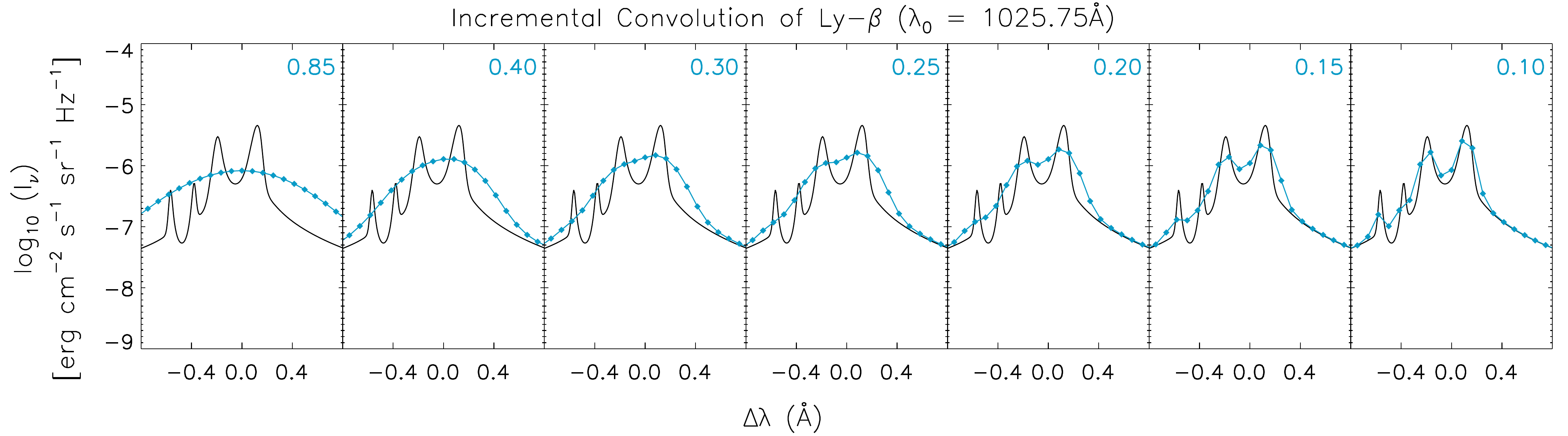}
	\caption{A snapshot of the Ly-$\beta$ line at t=18 s from the 3F10D8 simulation, convolved with Gaussians of increasingly narrow FWHM values. The FWHM of the Gaussian is stated in blue (in angstroms), while the resulting profile is rebinned to SPICE's wavelength spacing (0.083 \AA \ per pixel). \label{Figure19}}
\end{figure*}

In Figure \ref{Figure19}, snapshots of the Ly-$\beta$ line are shown at t=18 s for the 3F10D8 model after convolution with Gaussians of increasingly narrow FWHM values. The profiles are then rebinned to SPICE'S wavelength spacing (0.083 \AA \ per pixel). It can be seen that clear detection of the secondary line component and the central reversal in the line core would require an instrument with an instrumental profile around 0.2 \AA \ wide, while the central reversal is visible from FWHM values smaller than around 0.25 \AA.

\section{Discussion and conclusions} \label{sec:discussion}

The modelling of these chromospheric lines has led to some interesting results across a variety of different beam injection parameters. It should, however, be stressed that the RADYN simulations are not without approximation. Flares are not one-dimensional, single-loop structures. However, as a numerical experiment, the output from RADYN can be illuminating and used to obtain predictions. Further understanding could be gained from attempting a multi-threaded approach, by the addition of multiple beams over a spread of time with varying $\delta$ and E$_{c}$ values.

The simulations used in this paper cover four different beam-injection schemes ranging from moderate to high flux with a variety of deposition heights. In this section, we outline the key results before briefly revisiting each simulation in more detail. The key results are:

\begin{enumerate}
	\item According to the simulations, the Lyman lines can be blueshifted.
	\item The Lyman lines in the simulations often have a centrally reversed core, as a result of S$_{\nu}$ being large between the core and wing formation heights.
	\item EVE, and instruments with similar properties, may not be able to detect blueshifts in the Lyman lines if their cores are centrally reversed. In this case, the blueshifted core is not resolved and the line will present an overall red asymmetry.
	\item Accounting for non-equilibrium effects appears to be a more important factor in obtaining consistent line profiles between RADYN and RH, with the effects of PRD less significant (although not negligible).
\end{enumerate}

The F10D3 simulation typically reveals blueshifted centrally reversed cores, as a result of S$_{\nu}$ being greater at altitudes below the core formation height. The overall effect of this is to remove blue-wing irradiance from the line profiles, so that observationally the line could be interpreted as being redshifted. Figure \ref{Figure4} shows that degradation of the lines by an EVE-like instrument removes any delicate features, and so it is clear that care should be taken when trying to interpret the flow direction obtained from Doppler shifted EVE lines. In this simulation, despite the line cores clearly being blueshifted, convolution with the instrumental profile smear the asymmetric red peaks and blueshifted central reverals, giving the overall appearance of a redshifted line core. This leads to velocity profiles (Figure \ref{Figure6}) that would be easy to be interpreted as suggesting downflows during the beam injection.

The F10D8 simulation alters the deposition altitude of the electron beam, and leads to differences in the resulting velocity structure of the atmosphere as compared to when a low-$\delta$ beam is used. The atmosphere is upflowing, with a prominent velocity gradient at the leading edge of the upflow. As with the F10D3 simulation, the line cores are both centrally-reversed and strongly blueshifted, leading to perceived downflows in the synthetic velocity profiles. Compared to the $\delta=3$ case, the line core appears to form in a narrower region in altitude, and better samples the flow structure of the atmosphere.

The 3F10D8 simulation presents an interesting case of a secondary source of line emission. An additional line component is linked to a high-velocity atmospheric upflow. While the secondary feature is self-reversed, its overall effect is to enhance the blue wing and counteract the blueshifted absorption in the centrally reversed core. From Figure \ref{Figure3}, it can be seen that the velocity profiles throughout the heating stage are initially not too dissimilar from those in the F10D3 simulation, with differences arising due to the initiation of the secondary line source. It is interesting to note that despite the clear presence of a fast upflow in the line contribution functions between t=10-20 s (Figures \ref{Figure12}b \& c), its presence is not clearly detected in the Doppler velocity profiles because the weak emissive features in the lines are lost when degraded to EVE's resolution.

The F11D3 simulation presents an emitting feature during the relaxation process which appears to be linked to the flow structure in the atmosphere. Around 25 seconds after the energy deposition stops, an atmospheric flow propagates down from the corona and rebounds from a height very close to the transition region, transitioning back to flowing upwards. This height is also where the majority of the line core emission is being formed (Figures \ref{Figure15}c \&d), and from Figure \ref{Figure16} it appears that the rebounding flow causes a rise in density which results in collisional excitation of the levels in hydrogen. This leads to line emission, which is affected by the flow structure meaning that downflows are registered in the velocity profile.

These simulations suggest that the Lyman lines are sensitive to atmospheric upflows, and that optically thin emission can be produced in either red or blue wings depending on the circumstances, while the core emission is always optically thick. More importantly, the simulations have illustrated the complications that may arise when interpreting the flow direction from genuine observations of these lines.

By emulating the effects of instrumental response from the EVE instrument (used to measure Doppler shifts in the lines in \citet{Brown2016}), it is clear that one must be careful when it comes to assigning a direction to a flow observed by a detector with a wide instrumental profile. Our simulations show that an instrument such as EVE is unable to observe flows in the centrally reversed part of a line, as any self-reversal will not be retained after being affected by the instrumentation. In similarity to \citet{Kuridze2015}, we find that a blueshift in the centrally reversed core of the Ly-$\alpha$ line (and higher order Lyman lines) can lead to an observed red asymmetry, which masks the true direction of the flow. Similarly, the 3F10D8 simulation shows that in cases where there are flows indicated by the line profile, but the overall profile is symmetric, then the flows may be difficult to detect with the EVE instrument.

It also seems that the effects of assuming CRD on the line profiles may not be too severe while the electron beam heats the atmosphere. Figure \ref{Figure7} shows that during the relaxation stage of the F10D3 simulation, the ratio of core to wing emission is not comparable between RADYN and RH, and that even the RH profiles computed with CRD are significantly different from those in RADYN. In order to produce more accurate model profiles for the Lyman lines, it would be preferable to take non-equilibrium effects into account while also applying the PRD formalism. Non-equilibrium effects appear to be the most significant factor in causing the RH profiles to deviate from those in RADYN, but PRD remains important during the decay phase. PRD should also not be ignored during the heating phase, as it is likely highly dependent on the electron density.

These simulations have revealed a variety of interesting atmospheric features. From Figures \ref{Figure2}, \ref{Figure8}, \ref{Figure11} and \ref{Figure14}, it is clear that the deposition of energy alters the altitude of the transition region and leads to atmospheric upflows in all three simulations. The flow speed generally increases with the amount of energy deposited, with the F10 simulations displaying the weakest flows. 

Furthermore, it is clear that the magnitude of the atmospheric flows is not fully represented by the apparent Doppler velocities in the Lyman lines after convolution with the instrumental profile, which only provide a sample of the atmospheric velocity. The flow speed of the Lyman lines is typically found to be several tens of km s$^{-1}$ in these simulations, which is not implausible for these lines \citep{Brown2016}. However, the apparent Doppler velocities rarely match the atmospheric speed (from RADYN) over the line formation region and generally do not return the correct flow direction when the line profiles being centrally reversed. Interpreting the correct flow direction from observations requires knowledge of whether the line profile is centrally reversed. While both red and blue line shifts are observed in each of the Doppler velocity profiles, the initial atmospheric flows initiated in RADYN are typically upflowing.

This work has shown that, similar to the results in \citet{Brown2016}, the Doppler velocities of the Lyman lines during flare-type simulations suggest flow speeds of around 20-40 km s$^{-1}$. However, care should be taken when interpreting observations of these lines, as instrumental effects have the ability to misrepresent the detailed structure of the line profiles and the Doppler velocities may not fully reveal the maximum speed or direction of an atmospheric flow. Furthermore, time-integration performed by the instrumentation must be considered, as this represents an additional loss of information.


\section{Acknowledgements}
SAB is grateful for the support from an STFC research studentship. LF and NL are grateful for STFC funding under grant numbers ST/L000741/1 and ST/P000533/1. GSK is supported by an appointment to the NASA Postdoctoral Program at Goddard Space Flight Center, administered by USRA through a contract with NASA. JdlCR is supported by grants from the Swedish Research Council (2015-03994), the Swedish National Space Board (128/15) and the Swedish Civil Contingencies Agency (MSB). The authors are grateful to M. Carlsson and the F-CHROMA collaboration for the production and availability of a grid of RADYN simulations. The research leading to these results has received funding from the European Community's Seventh Framework Programme (FP7/2007-2013) under grant agreement no. 606862 (F-CHROMA), and from the Research Council of Norway through the Programme for Supercomputing.

\bibliographystyle{aasjournal}

\end{document}